\pdfoutput=1
\documentclass[acmsmall,screen,nonacm]{acmart}
\usepackage{booktabs}
\usepackage{amsmath}
\usepackage{graphicx}
\usepackage{xcolor}
\usepackage{multirow}
\usepackage{array}
\usepackage{enumitem}
\usepackage{float}
\setcopyright{none}
\hypersetup{pdfauthor={Louis Yiven Zhu}}
\newcommand{\srr}{\mathrm{SRR}}

\title{Silent Revision: Measuring Undisclosed Change in the Safety Frameworks of Frontier AI Developers}
\author{Louis Yiven Zhu}
\orcid{0009-0001-5579-0340}
\email{louis.zhu@oii.ox.ac.uk}
\affiliation{\institution{Oxford Internet Institute, University of Oxford}\city{Oxford}\country{United Kingdom}}

\begin{document}

\begin{abstract}
Frontier AI developers publish safety frameworks that commit them to evidencing whether their models are dangerous. The European Union and California now treat these documents as instruments of accountability, and both already impose duties on their revision. Neither requires the developer's account to identify each change to a commitment, so neither makes the revision legible to a reader of that account. We introduce the \emph{silent revision rate}, the share of material changes to a framework's commitments that the developer's published account does not identify, and we release the versioned, hash-pinned corpus needed to compute it. The corpus contains every public version of the safety frameworks of the twelve developers that have published one, together with each provider's changelog, redline or announcement. We trace 710 commitment instances across twelve consecutive version pairs, code them against a frozen codebook, and check the coding with three independent re-runs, a blind re-coding of every disputed row and a second coding of a stratified sample. Three findings follow. First, 64\% of material changes (95\% CI 59 to 69) are silent under a strict standard and 30\% under a lenient one, so the account identifies a third of all material change only at the level of a class of commitments. Second, the form of the account matters less than its granularity, since narrative and itemised accounts differ by eight points that eight pairs cannot separate from chance, whereas the rate falls with the number of discrete items a provider publishes, and the one redlined pair we traced is silent only once the redline is set aside. Third, 75\% of traced changes weaken or remove a commitment, and weakenings are silent more often than strengthenings, at 70\% against 52\%. The statutory remedy therefore exists and specifies the wrong artefact. A justification explains why a framework changed, an enumeration states what changed, and only the latter makes revision auditable. We argue that publication duties should carry an enumeration duty, which one provider already meets, voluntarily and incompletely.
\end{abstract}


\maketitle

\section{Introduction}
\label{sec:intro}

A standard that can be revised without anyone noticing is not a standard that anyone can be held to. Ananny and Crawford \citep{ananny2018seeing} argued that the transparency ideal confuses seeing with knowing, and frontier AI safety frameworks now illustrate the distinction. Every major developer of frontier AI models publishes a document stating how it will decide whether a model is too dangerous to train or release. These documents carry different names, among them Responsible Scaling Policy, Preparedness Framework and Frontier Safety Framework. Each defines capability thresholds, commits the developer to evaluations that establish whether a model has crossed one, and specifies the consequence when it has. In the vocabulary of the social sciences they are standards, because they render a contested notion of catastrophic risk commensurable as levels and procedures \citep{espeland1998commensuration,timmermans2010world}, and they function as regulatory science, because their authority rests on the visible propriety of the procedure that produces a judgement \citep{jasanoff1990fifth,porter1995trust}.

That authority has since acquired legal force in two jurisdictions. The European Union's General-Purpose AI Code of Practice requires signatories to maintain a safety and security framework and to provide the AI Office with each update \citep{eu2025gpai}. California's Transparency in Frontier Artificial Intelligence Act requires large developers to publish a frontier AI framework and, on any material modification, to publish the modified framework with a justification within thirty days \citep{california2025sb53}. Framework text is consequently the object that regulators, auditors and the public consult when they ask what a developer has promised, and a growing literature assesses it \citep{stelling2025evaluating,alaga2024grading,kasirzadeh2024measurement,pistillo2025towards,metr2024common,dsit2023emerging}.

Every one of those assessments scores a framework at a moment in time. Frameworks are revised, however, and both statutes are content with a snapshot of the revised document. The most thorough assessment to date observes that without changelogs an outsider cannot distinguish adaptation from weakening, and it itemises the changes between versions for two providers \citep{stelling2025evaluating}; it does not turn that observation into a measure, which is the measurement this paper supplies. The question sits at the decision end of evaluation science. We do not ask whether an evaluation measures what it claims; we ask whether a commitment to run it can be relied on once published.

We therefore put three questions to the corpus, in order of increasing specificity. \textbf{RQ1} concerns magnitude. When a developer revises its framework, what share of the material changes to its commitments can a reader identify from the developer's own account? \textbf{RQ2} concerns form. Does the kind of account a developer publishes, whether a redline, an itemised changelog, a narrative announcement or nothing, correspond to how much change stays silent? \textbf{RQ3} concerns direction. Do commitments tend to strengthen or weaken across revisions, and does the visibility of a change depend on its direction?

\begin{figure}[!t]
\centering
\includegraphics[width=\linewidth]{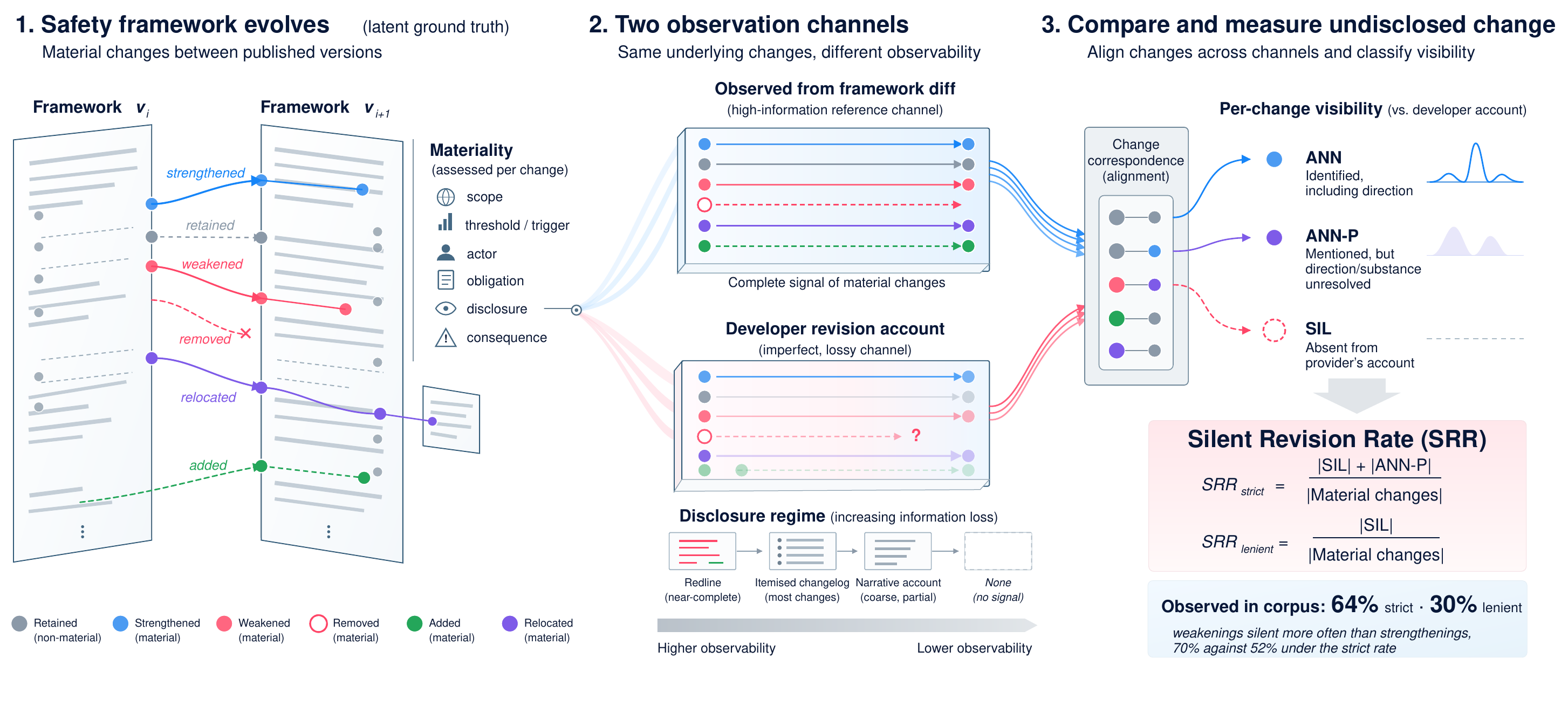}
\caption{The silent revision rate as a measurement construct. Material changes between versions (left) reach a reader through the text and through the developer's account, whose completeness varies by regime (centre); aligning the two classifies each change as announced, partially announced or silent (right). The pooled rates and the direction contrast at the lower right are developed within pairs in Section~\ref{sec:results}.}
\Description{Three-panel diagram. Left, two framework versions with commitments marked strengthened, weakened, removed, relocated or added. Centre, the same changes seen through the framework diff and through the developer's revision account, with four disclosure regimes ordered from redline to none. Right, each change classified as announced, partially announced or silent, the two rate definitions, and the corpus values of 64 percent strict and 32 percent lenient.}
\label{fig:pipeline}
\end{figure}

Answering these questions requires two things that did not previously exist, a corpus and a measure. We assemble every public version of the safety frameworks of the twelve developers that published one after the 2024 AI Seoul Summit, together with each developer's account of each revision. We then define the \emph{silent revision rate} as the share of material changes to a framework's commitments that the account does not identify. Figure~\ref{fig:pipeline} shows the construct, and a frozen codebook with two explicit thresholds operationalises it and exposes the judgement it embeds.

We argue that the auditability of a safety framework depends on the legibility of its revision, that legibility is measurable from the public record, and that in the current record it is low and non-random. We contribute (i) the corpus, comprising 43 labelled framework versions and silent same-label re-uploads, 9 companion documents and every revision account; (ii) the measure and codebook, with a granularity check on the denominator; (iii) first estimates with confidence intervals across twelve version pairs, three reliability checks of the coding, and the finding that weakenings are more often silent than strengthenings, pooled and within seven of eight pairs; and (iv) a compliance-gap argument, since both jurisdictions already impose a justification duty on revision while no form of account short of a redline identifies more than half of material change.

\section{Background}
\label{sec:background}

After the 2024 Seoul Summit, twelve developers published frameworks describing how they manage catastrophic risk from their most capable models, a development Anderljung et al.\ \citep{anderljung2023frontier} anticipated and Karnofsky \citep{karnofsky2024ifthen} set out in terms of if-then commitments. A literature quickly formed around the documents. The UK government \citep{dsit2023emerging} set out the safety processes such frameworks should contain and METR \citep{metr2024common} catalogued their common elements; Schuett et al.\ \citep{schuett2023towards} surveyed expert opinion; Alaga et al.\ \citep{alaga2024grading} proposed a grading rubric; Koessler et al.\ \citep{koessler2024risk} analysed threshold design; Kasirzadeh \citep{kasirzadeh2024measurement} identified six measurement challenges; and Pistillo \citep{pistillo2025towards} and Campos et al.\ \citep{campos2025frontier} argued for specificity and alignment with established risk management. The most thorough assessment, by Stelling et al.\ \citep{stelling2025evaluating}, scores twelve providers against 65 criteria. Every one of these instruments scores a snapshot; Stelling et al.\ report changes for the two providers that happened to revise during their window but do not carry longitudinal tracking out.

A second body of work concerns what developers disclose, and it establishes both the expectation of documentation and its limits. Model cards \citep{mitchell2019model} and datasheets \citep{gebru2021datasheets} set the template, Liang et al.\ \citep{liang2024systematic} showed across 32,111 model cards how unevenly it is filled, the Foundation Model Transparency Index \citep{bommasani2023fmti,bommasani2024fmti,wan2025fmti} finds developer transparency declining after an initial improvement, and Bommasani et al.\ \citep{bommasani2024reports} and Kolt et al.\ \citep{kolt2024responsible} set out what responsible reporting should contain. All of this measures whether a developer discloses a category of information, and none of it measures whether a disclosed document's later revision is itself disclosed, which is the second-order property this paper isolates. Mittelstadt \citep{mittelstadt2019principles} made the parallel point about ethics principles, which bind nobody until something converts them into enforceable practice.

That conversion is the concern of the auditing literature which supplies our institutional frame. Raji et al.\ \citep{raji2020closing,raji2022outsider} defined internal algorithmic audit and argued that it cannot substitute for third-party oversight, and Costanza-Chock et al.\ \citep{costanzachock2022who} recommended mandating the disclosure of key components of audit findings for peer review. Wachter et al.\ \citep{wachter2017why} showed how far a provision as written can sit from what a reader obtains under it. Our measure belongs to this tradition as an outsider-oversight instrument applied to the governing documents themselves, and our governance argument extends the disclosure recommendation from audit results to the revision of the standards audited against.

Whereas those bodies of work define the object, a third supplies the method. Jacobs and Wallach \citep{jacobs2021measurement} imported measurement modelling into the study of algorithmic systems, Wallach et al.\ \citep{wallach2025position} formalised the path from background concept to instrument, and Weidinger et al.\ \citep{weidinger2025toward} and R\"ottger et al.\ \citep{rottger2025safetyprompts,rottger2024political} showed what an evaluation science requires of its instruments; we treat the silent revision rate as such an instrument. The closest methodological precedent for our corpus lies outside AI governance, in Amos et al.'s longitudinal study of over a million privacy policies from the Internet Archive \citep{amos2021privacy}; we add the comparator the measure needs, namely the provider's own account of each change. The closest effort in subject matter is The Midas Project's AI Policy Watchtower \citep{midas2024watchtower}, a nonprofit monitor that logs edits to AI companies' policy documents, including edits the companies did not announce. It documents the phenomenon without defining a measure, reporting reliability or testing for asymmetry, and this paper is its first systematic measurement.

Finally, the social-scientific study of organisations supplies two candidate mechanisms, and we keep them distinct because the results discriminate between them. Vaughan \citep{vaughan1996challenger} explains the Challenger launch decision through structural secrecy, the routine consequence of specialisation that prevents any part of an organisation from seeing the aggregate of its own deviations. That mechanism predicts uneven silence, because the people who know which changes they intended write the changelog and everything outside their attention falls outside the record; it does not predict directional silence. Meyer and Rowan \citep{meyer1977institutionalized} and Brunsson \citep{brunsson1989organization} describe organisations that decouple the formal structure they display from the activity they conduct, and that mechanism does predict that unfavourable changes will be less visible than favourable ones.

\section{Corpus}
\label{sec:corpus}

We collected every publicly released version of the standing catastrophic-risk policy of each of the twelve developers that published one following the Seoul Summit (Appendix~\ref{apx:manifest} lists them). Alongside each version we collected the provider's own account of the revision, whether an in-document changelog, a version-history table, a published redline or the announcement post released with the version. Following Stelling et al.\ \citep{stelling2025evaluating}, we exclude system and model cards because they report point-in-time implementation.

Before retrieving anything, we established each provider's version list from primary sources, consulting framework pages and version histories first, then announcement posts, then Internet Archive capture histories. Secondary trackers located candidates but never served as sole evidence, and we never inferred a version from a numbering gap. We stored each file as published together with a plain-text extraction, a SHA-256 hash recomputed from disk and its provenance, and all nine passages that Stelling et al.\ quote with page references matched.

The resulting manifest has 52 rows, of which 43 are framework rows and 9 are companion documents. Thirty-five files came from providers, fourteen from the Internet Archive and one from a gated portal; three known versions could not be retrieved. Seven providers have at least two labelled versions and enter the drift analysis. Anthropic has nine versions, Google DeepMind four and xAI five, while OpenAI, Meta, Microsoft and Naver have two each. In addition, providers replaced nine files at the same URL or under the same version label with changed text and no new identifier, and we retain these as separate rows because a reader who downloaded the framework before and after would hold different documents bearing the same name (Appendix~\ref{apx:variants}).

Because a commitment that leaves a framework may reappear elsewhere, we also collected companion documents and consulted them to distinguish removal from relocation. One companion is itself a finding. Anthropic's Frontier Compliance Framework carries an itemised changelog describing four versions between December 2025 and July 2026, while the three earlier texts have been withdrawn from the portal that hosts them, so a reader can see what Anthropic says changed but cannot verify it. The same document commits to a changelog with justifications within thirty days of any material update, which tracks the statutory language of TFAIA almost verbatim (Appendix~\ref{apx:fcf}).

Across the nineteen consecutive labelled pairs, providers account for revision in four ways. A \emph{redline} is a full marked-up diff, published by Anthropic for every RSP revision from version 2.2 onward (five pairs). An \emph{itemised} account is a changelog listing individual changes (six pairs). A \emph{narrative} account is prose describing the revision, typically an announcement post (four pairs). \emph{None} means no account of any kind (four pairs, all xAI). We treat the four as an ordered typology of legibility, with the caveat that a redline shows every textual change without saying which are material or in which direction they move (Section~\ref{sec:limits}). Appendix~\ref{apx:pairs} assigns every pair. We release the corpus, manifest, codebook, coding sheets, adjudication notes and scripts, with our data under CC BY 4.0 and code under MIT, while provider documents remain under their owners' terms (\url{https://github.com/louisyzhu/frontier-safety-framework-corpus}, archived at \url{https://doi.org/10.5281/zenodo.22670700}). Every statistic is recomputed from the released coding sheet by the released script; the coding itself is an archived output of the procedure in Section~\ref{sec:method} and is reproducible only by re-running it (Appendix~\ref{apx:repro}).

\section{Method}
\label{sec:method}

The procedure is systematic content analysis in Krippendorff's sense \citep{krippendorff2019content}, applied to legally operative documents in the manner Hall and Wright \citep{hall2008systematic} set out for judicial opinions, and it takes its validity vocabulary from Adcock and Collier \citep{adcock2001measurement}. Krippendorff identifies unitising as the decision that most shapes a content analysis and is least visible in its results. Our unit is a \emph{commitment}, a statement in which the provider commits itself to a practice at any strength from \emph{must} to \emph{may}. Present-tense statements of practice count as commitments at the strongest rung, because in a policy document the descriptive present is the standing-commitment register. We code every commitment into one of nine categories (Appendix~\ref{apx:codebook}). Six are \emph{evidentiary}, because they concern how the provider will evidence risk or capability, namely scope of evaluation (EC1), trigger and threshold (EC2), method (EC3), third-party involvement (EC4), disclosure (EC5) and the consequence a result obligates (EC6). Three further strata, governance (G), security (S) and mitigation (M), are coded for completeness and reported separately.

For each consecutive version pair $(v_i, v_{i+1})$ we trace every commitment in $v_i$ to its counterpart in $v_{i+1}$, and we scan $v_{i+1}$ for commitments with no antecedent. Each traced commitment receives one of six outcomes, namely retained (R), strengthened (S), weakened (W), removed (X), relocated (L), which means it survives only in a companion document or a non-binding recommendations section, or added (A). When a commitment moves in both directions at once we code W by rule and flag it MIX. A change is \emph{material} if it alters at least one of six dimensions, namely scope, threshold or trigger, actor, obligation strength, disclosure scope or consequence. These six are one operationalisation of the systematised concept, chosen because each corresponds to a way the same commitment could bind differently, and Adcock and Collier's content-validation question, whether the indicators exhaust the concept, is answered in Appendix~\ref{apx:codebook} with the alternatives considered. Two rules follow drafting doctrine. The obligation ladder from \emph{must} through \emph{may} follows the mandatory and permissive distinction that Scalia and Garner \citep{scalia2012reading} catalogue, Their companion canon presumes that ``include'' introduces a nonexclusive list, and our enumeration rule is consistent with it, because a nonexclusive list still guarantees coverage of every item it names. We therefore treat an enumeration introduced by ``including'' as scope-defining, so that dropping a named item is material, whereas one introduced by ``for example'' is illustrative and the rule it illustrates stands without it. TFAIA uses ``material modification'' without defining it; our definition is the narrower one, since it applies to individual commitments, where the statute applies to the framework as a whole.

For each material change we then ask whether a reader of the provider's account of this revision, and nothing else, would learn that this commitment changed in this direction. We record the answer with four codes. A change is \emph{announced} (ANN) when the answer is yes, including where the account names a class of commitments and states the direction of travel. It is \emph{partially announced} (ANN-P) when the account names the commitment or its class but not the direction or substance. It is \emph{silent} (SIL) when an account exists and does not identify the change even at class level, and \emph{no changelog} (NCL) when the provider published no account. Only the provider's own publications count as an account.

Let $M_p$ denote the set of material changes on pair $p$, partitioned into announced $A_p$, partially announced $P_p$ and silent $S_p$. We define
\begin{equation}
\srr^{\mathrm{strict}}_p = \frac{|S_p| + |P_p|}{|M_p|}, \qquad
\srr^{\mathrm{lenient}}_p = \frac{|S_p|}{|M_p|}, \qquad
\srr^{\mathrm{strict}}_p - \srr^{\mathrm{lenient}}_p = \frac{|P_p|}{|M_p|} .
\label{eq:srr}
\end{equation}
The difference between the two forms is exactly the share of partially announced changes, and it exposes the judgement the measure embeds. The denominator is a granularity choice, since splitting one commitment into three raises silence mechanically when accounts describe change at class level, so we recompute the rate at section granularity (Appendix~\ref{apx:robust}). Every rate carries a Wilson score interval, which Brown, Cai and DasGupta \citep{brown2001interval} recommend over the Wald interval at the small $n$ of several pairs \citep{wilson1927probable}. The corpus is a census of the population, and we report intervals in the sense Berk, Western and Weiss \citep{berk1995statistical} give them for apparent populations, as statements about the process that generated the observed revisions. On a redlined pair every textual change is shown, so $\srr$ equals zero and we report those five pairs as regime-complete. To test the zero empirically we traced one redlined pair, RSP 3.0 to 3.1, in full. It contains 61 commitments and five material changes, and the redline marks every one of them; with the redline set aside, so that only the version-history entry and the in-document changelog count, two of the five are silent under the strict reading and one under the lenient (Appendix~\ref{apx:robust}).

For RQ2, changes cluster within pairs and a change-level Fisher test overstates precision \citep{cameron2015practitioner}, so our primary test is an exact permutation over the 56 assignments of the eight pair labels to three narrative and five itemised. Rank correlations between the rate and two features of the account, its length in words and its number of provider-published items, test whether verbosity or enumeration explains silence. For RQ3 we report the weakening share per pair and under leave-one-provider-out, since no theory predicts symmetric revision, and we compare silence between weakenings and strengthenings pooled and within each pair. All comparisons are descriptive associations within the corpus.

Coding then proceeded in two stages, a first pass and an adjudication of it. An agentic language-model system of the Claude family assisted us with the first pass, operating under codebook version 0.1 as its sole instruction with the two versions, the candidate-sentence list and the revision account as inputs. Version 0.2, frozen the next day, settled six questions that the first pass raised and introduced the partial-announcement code (Appendix~\ref{apx:codebook}). Pangakis et al.\ \citep{pangakis2023automated} show that the accuracy of such annotation varies by task and must be validated against human labels for each construct, and Gilardi et al.\ \citep{gilardi2023chatgpt} show that its zero-shot accuracy can exceed that of crowd workers against trained annotators' labels; we follow the first finding and do not presume the second. The pass produced 710 rows, each carrying verbatim text from both versions, a rationale, a confidence grade and, for material changes, a verbatim quotation from the account or a record that none was found. Every quoted field was verified against the corpus text (2{,}130 checks), which establishes quotation fidelity only. We then adjudicated 244 of the 710 rows individually from the passages, the account and the codebook, comprising every row touched by a post-first-pass clarification, every low-confidence material row and the fifty agreement-sample rows. Adjudication changed 4 outcome codes and 56 announcement codes, 46 of them from SIL to ANN-P. Because the adjudicator designed the codebook and holds the hypothesis, the direction of that drift matters, and it ran against the strict finding. We accepted the remaining 466 rows after a spot-check of 45 produced one change (Appendix~\ref{apx:adjudication}).

Reliability rests on three checks, each performed by model coders, since no human coded independently of the author (Section~\ref{sec:limits}). First, we re-ran the first pass three times on all 710 rows under version 0.2 with the same model and task text. Each run received the original commitment list without codes and had no access to any earlier code. The runs agree with one another at Krippendorff's $\alpha$ \citep{krippendorff2019content,hayes2007answering} of 0.92 to 0.94 for materiality, 0.90 to 0.91 for outcome and 0.86 for announcement status on all rows, and at 0.76 to 0.80 for announcement status on the rows where the three-way choice between announced, partially announced and silent arises. Against the adjudicated codes, however, the runs agreed on announcement status at only 0.69 to 0.71, because they applied the partial-announcement code far more widely than the adjudication had. We therefore re-examined every row on which any run disagreed with an adjudicated code, 249 rows in all. We read each again against the two texts, the account and the codebook under four readings fixed in advance (Appendix~\ref{apx:codebook}) and recorded a decision and a reason for each. Second, fresh model instances re-coded those 249 rows blind, with the documents, the account and the fixed readings but none of our codes. They agreed with the re-examined codes on 93 percent of materiality codes, 83 percent of outcome codes and 75 percent of announcement codes ($\alpha$ of 0.43, 0.78 and 0.61; the first is low because 93 percent of these rows are material and chance agreement is correspondingly high). We adjudicated each of the 92 rows carrying a disagreement against the texts and changed 37 of them. The two steps together changed 19 materiality codes, 43 outcome codes and 125 announcement codes relative to the adjudication, 84 of them from silent to partially announced (Appendix~\ref{apx:adjudication}). They raised agreement between the re-runs and the final codes to 0.84 to 0.85 on announcement status over all rows, which is the level at which the runs agree with each other, and to 0.70 to 0.71 on the three-way rows. Third, two Claude models other than the one used for the first pass and the re-runs, Claude Opus 5.5 and Claude Sonnet 5.5, each received only the coder package of fifty stratified units with the codebook and the accounts. They reached $\alpha$ of 0.94 and 1.00 for materiality and 0.92 for outcome. On announcement status they reached 0.59 and 0.60 on the 33 three-way units (95\% intervals 0.30 to 0.82 and 0.34 to 0.81), below the threshold of 0.667 that Krippendorff sets for tentative conclusions. The two agree with each other at 0.94, 0.90 and 0.79. The announcement disagreements fall almost entirely on the boundary between announced and partially announced. Raw agreement with our codes is 0.76 on the three-way choice and 0.94 on the lenient split between silent and pointed-at. On the same units the second coders' strict rates are 0.52 and 0.55 against 0.58 under our codes, and their weakening shares 0.78 and 0.77 against 0.85, so a change of coder moves the magnitudes by a few points and the direction of neither finding. Every number in this paper uses the final codes, and Appendix~\ref{apx:adjudication} reports the headline rates under the earlier codes as well. We release the sample, its instructions, the three re-run sheets, the blind re-coding and the script that computes $\alpha$.

\section{Results}
\label{sec:results}

\begin{figure}[!b]
\centering
\includegraphics[width=0.74\linewidth]{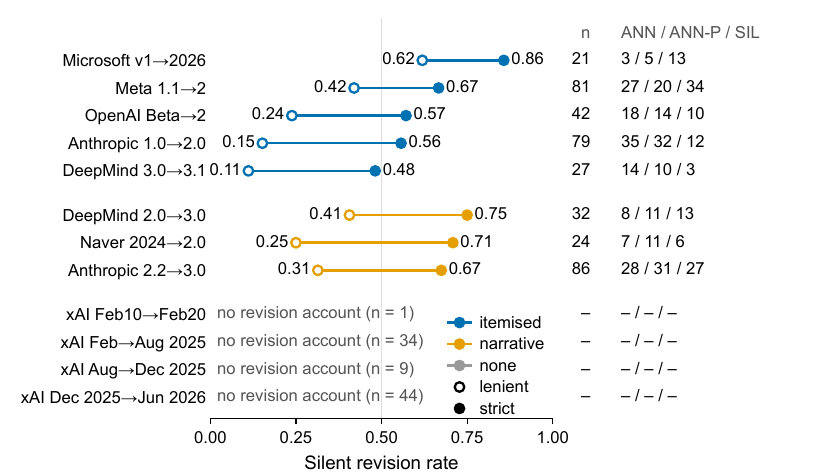}
\caption{Silent revision rate by pair, strict (filled) and lenient (open), grouped by disclosure regime. The right margin gives $n$ and the ANN, ANN-P and SIL counts; the xAI pairs published no account and have no defined rate.}
\Description{Dot plot of the strict and lenient silent revision rate for eight version pairs grouped as itemised or narrative, from 0.48 to 0.86 strict, with the four xAI pairs listed without a rate because no revision account exists.}
\label{fig:srr}
\end{figure}

For \textbf{RQ1}, most material change proves not to be identifiable in the provider's own account. Across the eight pairs with a revision account, 252 of 392 material changes are silent under the strict reading, a rate of 0.64 (95\% CI 0.59 to 0.69), and 118 of 392 under the lenient reading, a rate of 0.30 (0.26 to 0.35). The gap between the two readings is the share of partially announced changes, 134 of 392 or 0.34, and it is the central descriptive fact of the corpus. For a third of all material change, the account names the class of commitment that moved and leaves the reader to find the change. Seven of the eight accounted pairs exceed 0.55 on the strict reading and the eighth, DeepMind 3.0 to 3.1, stands at 0.48 (Table~\ref{tab:pairs}; Figure~\ref{fig:srr}). Evidentiary commitments run at 0.61 and 0.29, non-evidentiary strata at 0.71 and 0.32, and provider aggregates from 0.57 (OpenAI) to 0.86 (Microsoft) (Appendix~\ref{apx:category}). The rate depends on granularity, as it must. Collapsing to section level yields 170 units, of which 0.49 (0.42 to 0.57) are silent under the strict reading and 0.20 under the lenient one, and 0.58 under a majority rule within each section (Appendix~\ref{apx:robust}). The commitment-level rate is what a reader of one commitment experiences and the section-level rate is what a reader of the document experiences; under the strict reading both leave most change unaccounted for.

\begin{table}[t]
\centering\scriptsize\setlength{\tabcolsep}{3pt}
\caption{Twelve traced pairs. $n_{v_i}$ counts commitments traced from the earlier version; Mat.\ counts material changes including additions; W aggregates weakened, removed and relocated; $\srr_s$ and $\srr_l$ are the strict and lenient rates (intervals in Appendix~\ref{apx:stats}). Anthropic's five redlined pairs are regime-complete, so we did not trace them. Naver's 2024 version exists publicly only as an English summary page, so that pair over-counts additions.}
\label{tab:pairs}
\begin{tabular}{llrrrrrrrrrr}
\toprule
Pair & Regime & $n_{v_i}$ & Mat. & W & S & A & ANN & ANN-P & SIL & $\srr_s$ & $\srr_l$ \\
\midrule
Anthropic RSP 1.0\textrightarrow2.0 & itemised & 73 & 79 & 47 & 10 & 22 & 35 & 32 & 12 & 0.56 & 0.15 \\
Anthropic RSP 2.2\textrightarrow3.0 & narrative & 69 & 86 & 56 & 4 & 26 & 28 & 31 & 27 & 0.67 & 0.31 \\
OpenAI PF Beta\textrightarrow2 & itemised & 41 & 42 & 30 & 2 & 10 & 18 & 14 & 10 & 0.57 & 0.24 \\
DeepMind FSF 2.0\textrightarrow3.0 & narrative & 33 & 32 & 17 & 7 & 8 & 8 & 11 & 13 & 0.75 & 0.41 \\
DeepMind FSF 3.0\textrightarrow3.1 & itemised & 39 & 27 & 6 & 10 & 11 & 14 & 10 & 3 & 0.48 & 0.11 \\
Meta 1.1\textrightarrow2 & itemised & 49 & 81 & 11 & 13 & 57 & 27 & 20 & 34 & 0.67 & 0.42 \\
Microsoft v1\textrightarrow2026 & itemised & 40 & 21 & 9 & 4 & 8 & 3 & 5 & 13 & 0.86 & 0.62 \\
Naver 2024\textrightarrow2.0 & narrative & 17 & 24 & 11 & 4 & 9 & 7 & 11 & 6 & 0.71 & 0.25 \\
xAI draft Feb10\textrightarrow Feb20 & none & 42 & 1 & 1 & 0 & 0 & 0 & 0 & 0 & -- & -- \\
xAI Feb\textrightarrow Aug 2025 & none & 42 & 34 & 17 & 10 & 7 & 0 & 0 & 0 & -- & -- \\
xAI Aug\textrightarrow Dec 2025 & none & 43 & 9 & 1 & 3 & 5 & 0 & 0 & 0 & -- & -- \\
xAI Dec 2025\textrightarrow Jun 2026 & none & 48 & 44 & 24 & 10 & 10 & 0 & 0 & 0 & -- & -- \\
\bottomrule

\end{tabular}
\end{table}

For \textbf{RQ2}, the form of the account makes less difference than we expected, and its granularity more. Narrative pairs run at 0.70 (0.62 to 0.77) and itemised pairs at 0.61 (0.55 to 0.67) on the strict reading, a gap of eight points that no test separates from chance. The permutation test that respects nesting gives $p = 0.13$ for the pooled difference and $p = 0.18$ for the difference in per-pair means, and even the change-level Fisher test, which overstates precision, gives $p = 0.10$; on the lenient reading the two forms stand at 0.32 and 0.29. What does vary with the rate is enumeration. Across the eight pairs the rank correlation between the strict rate and the number of discrete items the provider published is $-0.58$, against $-0.24$ for the account's length in words (Appendix~\ref{apx:robust}). Eight pairs make neither a precise estimate, but Microsoft's six items cover 21 material changes at 0.86 and OpenAI's twelve cover 42 at 0.57. The redlined pair sharpens the point. On RSP 3.0 to 3.1 the redline marks all five material changes, whereas the version-history entry and the in-document changelog that the provider published alongside it leave two of the five silent (Appendix~\ref{apx:robust}). The gradient is also visible within a single provider. Anthropic alone publishes redlines, for five of its seven revisions, yet its largest revision from version 2.2 to 3.0 is one it did not redline. Of that revision's 86 material changes, 58 are silent under the strict reading and 27 even under the lenient one. The account explains at length why unilateral commitments were loosened and names classes of change without enumerating them. Among the changes it never identifies are the removal of a commitment to maintain a list of capabilities that may require stronger safeguards than the baseline, and the migration of its security-audit commitments into a compliance document that the account nowhere connects to them (Appendix~\ref{apx:examples}).

\begin{figure}[!htbp]
\centering
\includegraphics[width=0.9\linewidth]{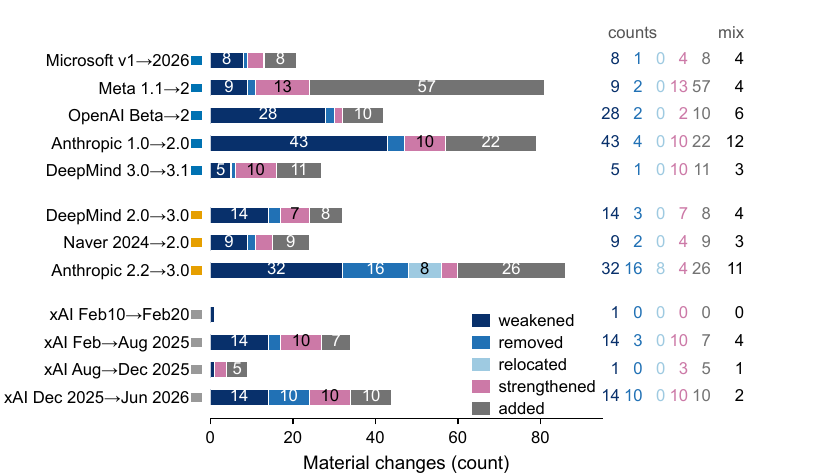}
\caption{Direction of material change by version pair, in the same row order as Figure~\ref{fig:srr}. Segment counts appear inside the segment where they fit and in the right-margin columns for every row; \emph{mix} counts the changes moving in both directions and coded weakened by rule, as flagged in the final codes (Appendix~\ref{apx:repro}). The tag at the left of each row gives the disclosure regime.}
\Description{Horizontal stacked bars for twelve version pairs giving counts of weakened, removed, relocated, strengthened and added material changes, with the counts and the number of mixed-direction changes listed in the right margin.}
\label{fig:direction}
\end{figure}

For \textbf{RQ3}, the direction of change is predominantly weakening, and weakening is more often silent than strengthening. Of 306 traced material changes across all twelve pairs, 230 weaken, remove or relocate a commitment, a share of 0.75 (0.70 to 0.80). Nine of the twelve pairs show a weakening majority (Figure~\ref{fig:direction}), the exceptions being DeepMind 3.0 to 3.1 (0.38), Meta (0.46) and xAI's four-change August to December 2025 pair, and dropping any one provider leaves the share between 0.67 and 0.78 (Appendix~\ref{apx:robust}). Excluding relocations or mixed changes leaves it at 0.75 and 0.70. A further 173 commitments are additions, concentrated in Meta's 2026 revision (57) and Anthropic's version 3.0 (26). Visibility depends on direction. Among changes with an account, weakenings are silent at 0.70 (130 of 187) and strengthenings at 0.52 (28 of 54), an odds ratio of 2.12 (Fisher's exact test, $p = 0.022$). Additions fall between at 0.62 (94 of 151); removals are the most silent outcome with more than eight cases at 0.74 (23 of 31), and six of the eight relocations are silent. Within pairs the asymmetry holds in seven of the eight with an account, by margins from 0.02 to 0.80. The exception is Anthropic's first revision, which runs the other way by 0.15 on ten strengthenings. Providers therefore announce the instruments they add more readily than the commitments they loosen, and the pattern is as visible across revisions as within them.

Two further observations strengthen the regulatory argument of Section~\ref{sec:discussion}. First, we split the pairs by the date of the later version around 1 January 2026, when TFAIA took effect. The three pairs closed before it run at 0.60 strict (92 of 153) and the five closed after at 0.67 (160 of 239), with weakening shares of 0.78 and 0.72. Three pairs a side support no causal claim, yet the justification duty coincided with no reduction in silence. Second, the trigger-and-threshold category (69 material changes) recurrently loses a quantitative anchor. DeepMind's R\&D threshold moved from ``substantially accelerating (e.g.\ 2x) from 2020--2024 rates'' to ``substantially accelerating from historical rates'' while the announcement reported that capability levels had been ``sharpened''. Third-party involvement is the smallest category (20 changes, 17 with an account, 0.59 silent). OpenAI's Beta commitment to ``continue to enable external research and government access'' survives in version 2 only as a discretionary clause on third-party evaluation, and none of its twelve changelog items mentions the change (Appendix~\ref{apx:examples}). Appendix~\ref{apx:variants} covers the nine same-label re-uploads, two of which are material.

\section{Discussion}
\label{sec:discussion}

We claimed that revision legibility is measurable from the public record and that in the current record it is low and non-random. The rate is computable for every pair with an account, and two-thirds of material change is silent on the strict reading and three in ten on the lenient. Of the two non-randomness claims, the direction result is established in the pooled data and holds within seven of eight pairs, whereas the regime result is not, since narrative and itemised accounts differ by eight points that eight pairs cannot separate from chance. What the data do show about accounts is where their information stops. One material change in three is identified at the level of a class, as when a changelog reports that safeguard requirements were made less prescriptive or that procedural commitments changed, and that pattern fits Vaughan's structural secrecy. A changelog written by the people who intended the changes records a revision as the organisation understood itself to be making it, theme by theme, and the individual commitments that moved under each theme fall outside its attention.

The direction result does not fit structural secrecy, and it fits decoupling. Structural secrecy predicts uneven silence yet gives no reason for silence to track the direction that reflects badly on the organisation. Meyer and Rowan's ceremonial conformity and Brunsson's organised hypocrisy do predict that pattern, since an organisation displaying a formal structure to outsiders describes what it adds and is quieter about what it loosens. We do not infer intent, since the people who write changelogs may simply champion the new instruments. Either way, visibility and content are not independent, so a reader who relies on the account alone sees without knowing \citep{ananny2018seeing}, and that gap is now a number.

Taken together, these findings identify a compliance gap in both regulatory regimes. TFAIA requires a large frontier developer that makes a material modification to publish the modified framework and a justification within thirty days \citep{california2025sb53}, and the EU Code requires signatories to assess their framework at least every twelve months, to accompany each update with a changelog stating how and why the framework changed, and to give the AI Office each update within five business days \citep{eu2025gpai}. Both regimes therefore already impose a duty on revision, and both specify an artefact that a class-level account satisfies. A justification explains why the framework changed, and a changelog that need only say how meets its duty with the kind of line that left a third of material change partially announced in our corpus. An enumeration states what changed, commitment by commitment and with direction. Our data show that no form of account short of a redline identifies more than half of material change, that length does not help, that enumeration helps only at the level of the individual change, and that silence did not fall after the duty took effect. Anthropic's compliance framework, whose Section 7.1 tracks the TFAIA language and then adds a changelog (Appendix~\ref{apx:fcf}), shows what drafting can produce one step past the statute. The same provider demonstrates that the remedy is voluntary, since it redlines five of seven revisions but not the largest and has withdrawn three compliance-framework versions whose changelog cannot be checked. The redlined pair we traced shows what the redline adds, because the two prose accounts published alongside it leave two of its five material changes silent and the redline leaves none. That record argues for a mandate, and we propose an enumeration duty. The developer should list each material change to a commitment, with its direction, in a form a reader can check against the text, since auditability is a property of how records are kept \citep{power1997audit}; a redline satisfies the duty mechanically.

The enumeration duty would also serve the audit ecosystem that the oversight literature describes. Outsider auditors check what a developer does against what it has said it will do \citep{raji2022outsider}, so an auditor who works from the current text alone inherits every silent change as the new baseline. Costanza-Chock et al.\ \citep{costanzachock2022who} ask that audit results be disclosed so that the ecosystem can check its own work, and the same logic applies one level up to the standard that audits measure against. The EU AI Office already receives each framework update within five business days \citep{eu2025gpai}, and it therefore holds the inputs to compute the rate for every signatory under the law as it stands. Researchers who score frameworks can adopt the same discipline at once by stating the version they scored and re-scoring on revision, since a score attached to a superseded text measures a promise the developer no longer makes.

\section{Limitations}
\label{sec:limits}

The most consequential limitation is that no human coded independently of the author. The three re-runs, the blind re-coding and the two second coders are all language models. Their agreement is therefore evidence about the procedure and about the transfer of the codebook between readers, while the validity of both rests on the argument of Section~\ref{sec:method}. Pangakis et al.\ \citep{pangakis2023automated} ask for validation against human labels, and we release the coder package so that a reader can supply it. Agreement on announcement status falls below Krippendorff's threshold on the three-way code, because readers draw the line between an announced and a partially announced change differently, although the pooled rates and the direction result survive a change of coder (Section~\ref{sec:method}). The second coders also applied the codebook that the adjudicator wrote, and the same adjudicator carried out the re-examination of disputed rows and the adjudication of the blind re-coding, so those steps test consistency with the codebook, while independence from the hypothesis remains untested.

The typology carries the next limitation at its top rung. Redlines score zero by construction, and the construction is partly circular, because a marked-up document shows every change without identifying which are material or in which direction they move. The traced redlined pair confirms the zero and also shows that a redline states no direction for two of its five changes. The regime is chosen by the provider, so the small and insignificant difference between narrative and itemised accounts may understate or overstate a real effect, and eight pairs cannot tell.

The remaining limitations bound what the statistics and the term silence can show. Changes nest within pairs, so we rely on the permutation test. The corpus has no comparison class, so we cannot say whether two-thirds is high against other regulated standards. Silence is evidence about external visibility only, and the term names an observable property of the record with no implication of motive. Finally, every claim about a named company is traceable to a hash-pinned text and a verbatim quotation, and we commit to a public errata policy and to versioning the corpus.

\section{Conclusion}
\label{sec:conclusion}

Frontier safety frameworks have become load-bearing in two legal regimes, so we built the corpus required to read them across time and defined a measure of revision legibility. A reader cannot identify two-thirds of material change from the provider's own account, and for a third of it the account names only the class of the change; the invisible share skews toward loosening, and only a redline closes the gap, which neither regime requires.

\section*{Ethical Considerations Statement}
This work analyses documents that developers published for public reading, and it involves no human participants. Every claim about a named developer rests on a hash-pinned document and a verbatim quotation, so each one can be checked against its source. We redistribute provider documents for research verification only and leave them under their owners' terms, while our own data and code carry open licences. The main adverse risk is that readers take silence for concealment. We therefore define silence as an observable property of the public record and state in Sections~\ref{sec:discussion} and \ref{sec:limits} that it carries no claim about intent. A second risk is that developers meet an enumeration duty in form alone, for instance with redlines that mark every edit and leave the material ones unflagged, and Section~\ref{sec:limits} names this weakness of the redline regime so that a regulator can guard against it. We keep a public errata file and a versioning policy, and we will correct any code that a developer shows to be wrong.

\section*{Acknowledgements}
The author thanks Emilio Barkett for early discussion of this project, and Paul R\"ottger, whose advice prompted the author to pursue empirical work in AI safety.

\section*{Generative AI Usage Statement}
The author conceived the study, defined the measure, wrote the codebook, collected and verified the corpus, produced the first coding decisions, and drafted and wrote the manuscript. We used Claude (Anthropic; \texttt{claude-fable-5-1} for the first pass, the re-runs and the blind re-coding; Claude Opus 5.5 and Claude Sonnet 5.5 as the two second coders) under our direction for four tasks. It assisted with the first-pass coding of the 710 commitment rows against the frozen codebook. It performed the three independent re-runs, the blind re-coding of the 249 disputed rows and the second coding of the fifty-unit sample, which Section~\ref{sec:method} reports as reliability checks. It checked bibliographic entries against their sources and generated the appendix tables and figure inputs from the released coding sheet by script. In a limited capacity, it also suggested grammar, style and formatting corrections to text we had written, each of which we accepted or rejected individually. The author reviewed every model output critically before use, adjudicated every disputed row against the texts before recording the final code, verified every quotation and reference, and recorded where model assistance entered the technical work in the released run log and coding notes. The author takes full responsibility for the originality, accuracy and integrity of the manuscript.

\bibliographystyle{ACM-Reference-Format}
\bibliography{refs}

\newpage
\appendix
\section*{Appendices}
Each appendix names the section and claim of the main text it supports. Appendix~\ref{apx:glossary} defines every term, abbreviation and acronym. Appendices~\ref{apx:codebook} and \ref{apx:calibration} reproduce the frozen codebook rules and the calibration examples that anchored coding (Section~\ref{sec:method}). Appendix~\ref{apx:stats} derives the statistics and Appendix~\ref{apx:robust} reports the robustness analyses (Sections~\ref{sec:method} and \ref{sec:results}). Appendix~\ref{apx:examples} quotes both versions of every change discussed in Sections~\ref{sec:results} and \ref{sec:discussion}. Appendix~\ref{apx:adjudication} tabulates the coding procedure. Appendix~\ref{apx:category} gives further results, Appendices~\ref{apx:pairs} to \ref{apx:fcf} tabulate the corpus described in Section~\ref{sec:corpus}, and Appendices~\ref{apx:repro} and \ref{apx:manifest} state what the release contains and how each number can be recomputed.

\section{Glossary of terms, abbreviations and acronyms}
\label{apx:glossary}
Table~\ref{tab:glossary} defines the terms the paper uses in a technical sense and expands every abbreviation; Section~\ref{sec:method} introduces each in context.
\begin{table}[H]\centering\small
\caption{Terms, abbreviations and acronyms used in the paper.}
\label{tab:glossary}
\begin{tabular}{lp{9.4cm}}
\toprule Term & Meaning \\ \midrule
Commitment & A statement in which the provider commits itself to a practice at any strength; the unit of analysis \\
Material change & A change to a commitment that alters scope, threshold or trigger, actor, obligation strength, disclosure scope or consequence \\
Revision account & The provider's own published statement of what changed in a revision (changelog, version table, redline or announcement post) \\
Disclosure regime & The form of the revision account: redline, itemised, narrative or none \\
Silent revision rate ($\srr$) & The share of material changes the revision account does not identify (Equation~\ref{eq:srr}); strict counts partial announcement as silent, lenient does not \\
Companion document & A document the provider publishes alongside the framework, consulted to distinguish removal from relocation \\
Silent same-label re-upload & A file replaced at the same URL or version label with changed text and no new identifier \\
Evidentiary category & A commitment category concerning how risk or capability will be evidenced (EC1 to EC6) \\
ANN / ANN-P / SIL / NCL & Announced / partially announced / silent / no changelog (announcement codes) \\
R / S / W / X / L / A & Retained / strengthened / weakened / removed / relocated / added (outcome codes) \\
MIX & Flag for a commitment that both strengthens and weakens; coded W by rule \\
EC1 to EC6 & Scope of evaluation, trigger and threshold, method, third-party involvement, disclosure, consequence \\
G / S / M & Governance / security / mitigation strata (non-evidentiary) \\
RSP, PF, FSF, RMF & Responsible Scaling Policy (Anthropic), Preparedness Framework (OpenAI), Frontier Safety Framework (DeepMind), Risk Management Framework (xAI) \\
FCF & Frontier Compliance Framework (Anthropic companion document) \\
CCL & Critical capability level (DeepMind's threshold term) \\
ASL & AI Safety Level (Anthropic's threshold term) \\
TFAIA & California Transparency in Frontier Artificial Intelligence Act (SB 53) \\
EU Code & General-Purpose AI Code of Practice under the EU AI Act \\
CI & 95\% Wilson score confidence interval \\
\bottomrule
\end{tabular}
\end{table}

\section{Codebook v0.2 (frozen 3 September 2026)}
\label{apx:codebook}
This appendix reproduces the rules that Section~\ref{sec:method} summarises. The adjudicator and the second coders applied these rules and nothing else. The first pass applied version 0.1, which lacked the clarifications in the final subsection, and the full document with its change history is in the release.
\subsection{Inclusion}
A row is a commitment if the provider is the grammatical subject (we, our, the company name, a named internal role) and the sentence carries a commitment verb or modal (will, must, commit, shall, intend, aim, expect, may, could, consider, plan) or states a practice in the descriptive present. Conditional commitments (``if X, we will Y'') are included. Statements describing the risk landscape, defining terms, describing other parties' obligations, or reporting a past point-in-time evaluation are excluded.
\subsection{Categories}
EC1 scope of evaluation (which capabilities or risk domains will be evaluated); EC2 trigger and threshold (when evaluations must occur and what result constitutes crossing); EC3 method (how evaluations are conducted); EC4 third-party involvement (external evaluation, government access, audits); EC5 disclosure (what evaluation methods and results will be published, to whom, when); EC6 consequence linkage (what action a specified result obligates). G governance and oversight; S security and containment; M mitigation and deployment measures not expressed as a consequence of an evaluation result.
\subsection{Tracing}
For each pair, start from the commitment list of $v_i$ and locate each commitment's counterpart in $v_{i+1}$, which is the passage governing the same category and the same object. Renumbering, relocation to a different section and rewording do not break correspondence. Relocation to a companion document or to a recommendations section is its own outcome (L). Then scan $v_{i+1}$ for commitments with no counterpart (A).
\subsection{Outcomes}
R retained (no material change); S strengthened (at least one materiality dimension moves toward greater obligation, broader scope, lower threshold, more third-party involvement or more disclosure); W weakened (at least one dimension moves the other way); X removed; L relocated; A added. Both directions present in one commitment yields W and the MIX flag.
\subsection{Materiality}
A change is material if it alters scope; threshold or trigger; actor (including addition or removal of a third party); obligation strength on the scale \emph{must / will / commit to / present-tense practice} $>$ \emph{intend to / aim to / expect to} $>$ \emph{may / could / consider} $>$ \emph{recommend / encourage}; disclosure scope; or consequence. Not material are rewording at the same force, reorganisation, typographical changes, updated cross-references, and changes to illustrative examples that do not alter the rule. Broadening a suspension condition is a change to consequence. Enumerations introduced by ``including'' or forming the operative content are scope-defining; enumerations introduced by ``e.g.'', ``for example'' or ``such as'' are illustrative.
\subsection{Alternatives considered for the materiality construct}
Three alternative operationalisations were considered and rejected before coding. Counting every textual edit (as a redline does) makes no distinction between rewording and change of obligation and would inflate the denominator with editorial changes. Scoring against an external rubric of framework quality (for instance the 65 criteria of Stelling et al.) measures the level of a framework and not the change in an individual commitment, and it would leave changes to commitments outside the rubric uncounted. Coding obligation strength alone (the modal ladder) would miss changes of scope, actor and consequence, which account for most of the material changes in the corpus. The six dimensions were chosen because each names a distinct way in which the same commitment can bind differently, and the released coding sheet records which dimension each material change engaged so that the choice can be revisited.
\subsection{Announcement status}
Compare each material change against the provider's own account of the revision, which comprises an in-document changelog or version table, prose in the document describing what changed, the provider's announcement post, or a provider-published redline. ANN when a reader of the account alone would know that this commitment changed in this direction, including class-level mentions that state direction; ANN-P when the account names the commitment or its class without direction or substance; SIL when the account exists and does not identify the change even at class level, with generic lines not counting; NCL when no account exists. Commentary by individuals, including employees in a personal capacity, is not a provider account.
\subsection{Clarifications adopted after the first pass}
(1) Present-tense practice statements are commitments at the top rung. (2) Enumerations follow the scope-defining and illustrative rule above. (3) Both-direction changes are W with MIX. (4) Class-level changelog lines without direction are ANN-P. (5) A commitment that leaves the RSP at version 3.0 and appears in a companion in the corpus is L, with the temporal caveat noted; OpenAI's Frontier Governance Framework post-dates PF version 2 by thirteen months and is not consulted for that pair. (6) In xAI's February to August 2025 revision, the ``may also provide'' sentence is a disclosure commitment (EC5) and its own row; the external red-team testing commitment traces to a different sentence and is W on the actor dimension.

\subsection{Rules applied in the re-examination}
The re-examination of the 249 disputed rows (Section~\ref{sec:method}), the blind re-coding of the same rows and the review that reconciled the two applied the rules above with four readings fixed in advance and recorded against each row. First, a class-level mention counts only when the class is specific enough to locate the affected commitments in the document, such as the ASL-3 safeguard requirements, the actors in scope of a security standard, the Risk Report provisions or a named section of procedural commitments; overview sentences such as ``key improvements include refined processes'' and generic lines do not count. Second, a located class whose stated direction applies to the change is ANN, under the sentence in section 6 that counts class-level mentions which state direction; a located class whose stated direction does not apply to the change, or that states none, is ANN-P. Third, a commitment that leaves the framework and survives in a companion document in the corpus, or in the framework's own recommendations section, is relocated (L), and its announcement status follows the account's treatment of that move. Fourth, enumerations are read by rule 11.2 as written, so a list introduced by ``including'' is scope-defining and a list introduced by ``such as'' or ``example'' is illustrative, whichever way the resulting materiality decision cuts.

\section{Calibration examples}
\label{apx:calibration}
The following cases anchored the coding described in Section~\ref{sec:method}. They are drawn from the corpus and from Stelling et al.'s Tables 9 and 10 \citep{stelling2025evaluating}, and each was verified against the primary documents.
\begin{itemize}[leftmargin=1.2em]
\item \textbf{A, announced weakening (ANN, W).} Anthropic RSP 2.2 to 3.0 removed unilateral pause commitments and separated company commitments from industry recommendations; the accompanying account explains the change.
\item \textbf{B, silent weakening (SIL, W).} OpenAI Preparedness Framework Beta to version 2, footnote 6, provides that models distilled, fine-tuned or quantised from a model below a High threshold will ordinarily not require additional safety measures; the twelve-item changelog does not mention it.
\item \textbf{C, announced consequence change (ANN, W).} OpenAI version 2, Section 4.3, allows safeguards to be adjusted if another developer releases a High or Critical system without comparable safeguards; changelog item 11 states it.
\item \textbf{D, weakening with no account (NCL, W).} xAI February to August 2025 replaced its commitment to external red-team testing of safeguards; no account exists.
\item \textbf{E, not material (R).} Renumbered capability levels with identical modal force and scope.
\item \textbf{F, de-commensuration of a threshold (W, EC2).} DeepMind FSF 2.0 to 3.0 replaced ``substantially accelerating (e.g.\ 2x) from 2020--2024 rates'' with ``substantially accelerating from historical rates''.
\item \textbf{G, loss of specificity (W, EC3).} FSF 2.0 listed what post-market monitoring draws on; FSF 3.0 says ``post-market monitoring''.
\item \textbf{H, modal drift (W, EC6).} ``The safety case will be updated through red-teaming'' became safety cases ``may be updated if deemed necessary''.
\item \textbf{I, marginal-risk provision (W, EC6).} FSF 3.0 allows marginal risk relative to competitors to inform deployment decisions; FSF 2.0 had no such provision.
\item \textbf{J, governance de-naming (W, G).} Three named councils became ``appropriate governance function''.
\item \textbf{K, announced removal with rationale (ANN).} Anthropic's version 3.0 account explains the removal of unilateral pause commitments at length.
\end{itemize}

\section{Statistical derivations}
\label{apx:stats}
This appendix supports the intervals and tests reported in Section~\ref{sec:results} and specified in Section~\ref{sec:method}.
\paragraph{Wilson score interval.} For $k$ silent changes among $n$ material changes and $z = 1.96$, the interval for the rate $p$ is
\begin{equation}
\frac{1}{1+z^2/n}\left[\hat p + \frac{z^2}{2n} \pm z\sqrt{\frac{\hat p(1-\hat p)}{n} + \frac{z^2}{4n^2}}\right], \qquad \hat p = k/n,
\end{equation}
which unlike the Wald interval remains inside $[0,1]$; Brown, Cai and DasGupta \citep{brown2001interval} recommend it for small $n$ on coverage grounds \citep{wilson1927probable}.
\paragraph{Per-pair intervals.} Strict rates with 95\% intervals are Anthropic 1.0 to 2.0, 0.56 (0.45, 0.66); Anthropic 2.2 to 3.0, 0.67 (0.57, 0.76); OpenAI, 0.57 (0.42, 0.71); DeepMind 2.0 to 3.0, 0.75 (0.58, 0.87); DeepMind 3.0 to 3.1, 0.48 (0.31, 0.66); Meta, 0.67 (0.56, 0.76); Microsoft, 0.86 (0.65, 0.95); Naver, 0.71 (0.51, 0.85). Lenient rates are Anthropic 1.0 to 2.0, 0.15 (0.09, 0.25); Anthropic 2.2 to 3.0, 0.31 (0.23, 0.42); OpenAI, 0.24 (0.13, 0.39); DeepMind 2.0 to 3.0, 0.41 (0.26, 0.58); DeepMind 3.0 to 3.1, 0.11 (0.04, 0.28); Meta, 0.42 (0.32, 0.53); Microsoft, 0.62 (0.41, 0.79); Naver, 0.25 (0.12, 0.45).
\paragraph{Regime comparison (RQ2).} The exact permutation test is the primary test. It assigns the narrative label to each of the $\binom{8}{3} = 56$ subsets of three pairs, recomputes the pooled strict difference, and reports the share of assignments at or above the observed difference of 0.085; that share is $0.13$. On the difference in per-pair means (observed 0.084) the share is $0.18$. For reference, pooled strict silence is 153 of 250 on itemised pairs and 99 of 142 on narrative pairs, and Fisher's exact test on the $2 \times 2$ table gives an odds ratio of 1.46 and $p = 0.10$; on the lenient reading the counts are 72 of 250 and 46 of 142, odds ratio 1.18, $p = 0.49$. The change-level test overstates precision because changes cluster within pairs \citep{cameron2015practitioner}. Under the codes that preceded the re-examination (Appendix~\ref{apx:adjudication}) the same tests gave $p = 0.071$, $p = 0.089$ and $p = 0.041$, so the regime difference is a casualty of the re-examination and we report it as not established.
\paragraph{Direction (RQ3).} Among 306 traced material changes, 230 are W, X or L, a share of 0.75 with Wilson interval (0.70, 0.80). Excluding the 54 MIX-flagged changes gives 177 of 252, share 0.70 (0.64, 0.76); excluding the eight relocations gives 222 of 298, share 0.75. Per-pair and leave-one-provider-out shares appear in Appendix~\ref{apx:robust}. Among changes with an account, strict silence is 130 of 187 for weakenings, 28 of 54 for strengthenings and 94 of 151 for additions; Fisher's exact test on weakenings against strengthenings gives an odds ratio of 2.12 and $p = 0.022$. By outcome, strict silence is 101 of 148 for W, 23 of 31 for X, 6 of 8 for L, 28 of 54 for S and 94 of 151 for A. On the lenient reading weakenings are silent at 61 of 187 and strengthenings at 16 of 54, odds ratio 1.15, $p = 0.74$, so the asymmetry lives in partial announcement.
\paragraph{Multiple comparisons.} Appendix~\ref{apx:category} reports nine category rates with intervals and no test; the paper draws no inference from differences between categories.
\paragraph{Reliability.} For two coders and nominal data, Krippendorff's $\alpha = 1 - D_o / D_e$, where $D_o$ is the observed disagreement across the coded units and $D_e$ the disagreement expected by chance from the marginal distribution of values \citep{krippendorff2019content}. The released script computes $\alpha$ separately for materiality, outcome and announcement status, with a bootstrap 95\% interval over units, and the final code of each row follows the precedence stated in Appendix~\ref{apx:repro}. Announcement status is compared on all 710 rows, where non-material rows carry NA and the xAI rows NCL, and on the three-way rows, where the final code is ANN, ANN-P or SIL; the first figure is inflated by rows on which no three-way choice arises, so the second is the one to weigh. Between the three independent re-runs of the first pass, $\alpha$ is 0.93, 0.94 and 0.92 for materiality, 0.90, 0.91 and 0.91 for outcome, and 0.86, 0.86 and 0.86 for announcement status on all rows, with bootstrap intervals no wider than 0.06, and 0.77, 0.76 and 0.80 for announcement status on the three-way rows. Against the final codes the three runs give 0.95, 0.94 and 0.94 for materiality, 0.91, 0.92 and 0.92 for outcome, and 0.84, 0.85 and 0.85 for announcement status on all rows and 0.70, 0.71 and 0.71 on the three-way rows; against the codes that preceded the re-examination they gave 0.70, 0.69 and 0.71 for announcement status on all rows. The blind re-coding of the 249 disputed rows, compared with the re-examined codes before the review that followed it, gives 0.43 (0.18, 0.64) for materiality, 0.78 (0.71, 0.83) for outcome and 0.61 (0.51, 0.70) for announcement status on the 211 three-way rows, with raw agreement 0.93, 0.83 and 0.75. The two second coders, Claude Opus 5.5 and Claude Sonnet 5.5, give 0.94 (0.77, 1.00) and 1.00 for materiality, 0.92 (0.82, 1.00) and 0.92 (0.82, 1.00) for outcome, and 0.59 (0.30, 0.82) and 0.60 (0.34, 0.81) for announcement status on the 33 three-way units of the fifty-unit sample; between themselves they give 0.94, 0.90 and 0.79. Collapsing announcement status to the strict split (announced against the rest) gives 0.64 and 0.58 with raw agreement 0.82 and 0.79, and to the lenient split (silent against the rest) 0.64 and 0.72 with raw agreement 0.94 and 0.94.

\section{Robustness analyses}
\label{apx:robust}
This appendix supports the granularity, account-length, per-pair direction and within-pair asymmetry results reported in Section~\ref{sec:results}.
\paragraph{Granularity.} Collapsing material changes to the section of $v_i$ in which a traced change occurs, or the section of $v_{i+1}$ in which an addition appears, yields 170 section-level units across the eight pairs with an account. A unit counts as announced if any change in it is ANN and as partially announced if any is ANN-P and none is ANN. The strict rate at this granularity is 0.49 (0.42, 0.57) and the lenient rate 0.20 (0.15, 0.27). Under a majority rule, in which a unit is silent if more than half its changes are, the strict figure is 0.58.
\paragraph{Account length and enumeration.} Table~\ref{tab:robust} gives, for each pair, the length of the revision account in words, the number of discrete items the provider published in it (bulleted, numbered or labelled entries, counted by hand from the account), the number of material changes with an account, the number the account identifies (ANN), and the strict rate. Across the eight pairs the Spearman correlation between the strict rate and words is $-0.24$, between the rate and words per material change $0.10$, between the rate and items $-0.58$, and between the rate and items per material change $-0.13$. The count of identified changes (ANN) is not used as a predictor because the rate is defined as its complement.
\begin{table}[H]\centering\scriptsize
\caption{Per-pair robustness quantities. Words and Items describe the revision account; Mat.\ (acct) and ANN count material changes on pairs with an account and those the account identifies; Traced counts material changes excluding additions, over which Weak.\ share is computed; the last column gives strict silence among weakenings and among strengthenings, with counts.}
\label{tab:robust}
\resizebox{\linewidth}{!}{\begin{tabular}{llrrrrrrrr}\toprule Pair & Regime & Words & Items & Mat.\ (acct) & ANN & $\srr_s$ & Traced & Weak.\ share & Silent W / S \\ \midrule
Anthropic 1.0\textrightarrow2.0 & itemised & 4710 & 11 & 79 & 35 & 0.56 & 57 & 0.82 & 0.45 / 0.60 (47/10) \\
Anthropic 2.2\textrightarrow3.0 & narrative & 14274 & 3 & 86 & 28 & 0.67 & 60 & 0.93 & 0.77 / 0.75 (56/4) \\
OpenAI Beta\textrightarrow2 & itemised & 1971 & 12 & 42 & 18 & 0.57 & 32 & 0.94 & 0.63 / 0.50 (30/2) \\
DeepMind 2.0\textrightarrow3.0 & narrative & 863 & 3 & 32 & 8 & 0.75 & 23 & 0.74 & 1.00 / 0.57 (17/7) \\
DeepMind 3.0\textrightarrow3.1 & itemised & 358 & 6 & 27 & 14 & 0.48 & 16 & 0.38 & 1.00 / 0.20 (6/10) \\
Meta 1.1\textrightarrow2 & itemised & 1353 & 7 & 81 & 27 & 0.67 & 24 & 0.46 & 0.55 / 0.46 (11/13) \\
Microsoft v1\textrightarrow2026 & itemised & 291 & 6 & 21 & 3 & 0.86 & 13 & 0.69 & 0.89 / 0.75 (9/4) \\
Naver 2024\textrightarrow2.0 & narrative & 2139 & 3 & 24 & 7 & 0.71 & 15 & 0.73 & 0.91 / 0.75 (11/4) \\
xAI Feb10\textrightarrow Feb20 & none & -- & -- & -- & -- & -- & 1 & 1.00 & -- \\
xAI Feb\textrightarrow Aug 2025 & none & -- & -- & -- & -- & -- & 27 & 0.63 & -- \\
xAI Aug\textrightarrow Dec 2025 & none & -- & -- & -- & -- & -- & 4 & 0.25 & -- \\
xAI Dec 2025\textrightarrow Jun 2026 & none & -- & -- & -- & -- & -- & 34 & 0.71 & -- \\
\bottomrule\end{tabular}
}
\end{table}
\paragraph{Before and after TFAIA.} Splitting the eight accounted pairs by the date of the later version around 1 January 2026, the three pairs closed before (Anthropic 1.0 to 2.0, OpenAI, DeepMind 2.0 to 3.0) run at 0.60 strict (92 of 153) and 0.23 lenient, and the five closed after run at 0.67 strict (160 of 239) and 0.35 lenient. Across all twelve pairs the weakening share is 0.78 (113 of 144) before and 0.72 (117 of 162) after.
\paragraph{Direction by pair and by provider.} Table~\ref{tab:robust} also gives the weakening share for each of the twelve pairs and, for the eight pairs with an account, the strict silence of weakenings and of strengthenings separately. Nine of twelve pairs show a weakening majority. Table~\ref{tab:loo} gives the pooled weakening share after dropping each provider in turn; it lies between 0.67 and 0.78.
\begin{table}[H]\centering\small
\caption{Leave-one-provider-out weakening share among traced material changes.}
\label{tab:loo}
\begin{tabular}{lrrr}\toprule Provider dropped & Weakening & Traced & Share \\ \midrule
Anthropic & 127 & 189 & 0.67 \\
Google DeepMind & 207 & 267 & 0.78 \\
Meta & 219 & 282 & 0.78 \\
Microsoft & 221 & 293 & 0.75 \\
Naver & 219 & 291 & 0.75 \\
OpenAI & 200 & 274 & 0.73 \\
xAI & 187 & 240 & 0.78 \\
\bottomrule\end{tabular}

\end{table}
\paragraph{The redlined pair.} Tracing RSP 3.0 to 3.1 from scratch under the same procedure yields 61 commitments, of which 56 are retained, four weakened and one added, so five material changes, four of them evidentiary. The redline marks all five. Counting all three provider publications as the account, the strict and lenient rates are both 0 of 5. Counting only the version-history entry and the in-document changelog, the strict rate is 2 of 5 and the lenient rate 1 of 5. The changelog names the class of one change (how internal feedback on Risk Reports is used) without its direction, and neither prose account mentions the other (the comparison class that brings an internal model into the scope of a Risk Report). No provider text states the direction of those two changes, so a redline reader learns that they happened and not which way they moved. The redline comprises 15 mechanical hunks, of which six are editorial (version number, date, two one-word insertions, the changelog entry and one permissive sentence) and nine produce the five changes.

\section{Verbatim text for every change discussed in the main text}
\label{apx:examples}
Each entry gives the commitment identifier from the released tracing sheet, the earlier and later text as extracted, the provider's account where one exists, and every adjudication note, including the note from the re-examination where the row was disputed. Sections~\ref{sec:results} and \ref{sec:discussion} cite these entries, and the remaining entries illustrate the codes that the calibration examples of Appendix~\ref{apx:calibration} anchor.
\sloppy
\paragraph{ANT-1-001 (EC6; W; SIL).} \emph{Anthropic v1.0 \textrightarrow\ v2.0.}\\
\textbf{Earlier:} Anthropic's commitment to follow the ASL scheme thus implies that we commit to pause the scaling2 and/or delay the deployment of new models whenever our scaling ability outstrips our ability to comply with the safety procedures for the corresponding ASL.\\
\textbf{Later:} In any scenario where we determine that a model requires ASL-3 Required Safeguards but we are unable to implement them immediately, we will act promptly to reduce interim risk to acceptable levels until the ASL-3 Required Safeguards are in place: ... Interim measures: The CEO and Responsible Scaling Officer may approve the use of interim measures that provide the same level of assurance as the relevant ASL-3 Standard but are faster or simpler to implement.\\
\textbf{Provider's account:} none found\\
\textbf{Adjudication:} Agree; pause commitment becomes "act promptly to reduce interim risk". Headline example.

\paragraph{ANT-2-014 (EC1; X; SIL).} \emph{Anthropic v2.2 \textrightarrow\ v3.0.}\\
\textbf{Earlier:} We will also maintain a list of capabilities that we think require significant investigation and may require stronger safeguards than ASL-2 provides.This groupof capabilities could pose serious risks, but the exact Capability Threshold and the Required Safeguards are not clear at present.\\
\textbf{Later:} NONE\\
\textbf{Provider's account:} none found\\
\textbf{Adjudication:} Accepted from first pass (spot-check tier). Re-examination: Confirmed.

\paragraph{ANT-2-038 (EC4; L; SIL).} \emph{Anthropic v2.2 \textrightarrow\ v3.0.}\\
\textbf{Earlier:} Audits:Develop plans to (1) audit and assess thedesign and implementation of the security program and (2) share these findings (and updates on any remediation efforts) with management on an appropriate cadence. We expect this to include independent validation of threat modeling and risk assessment results; a sampling-based audit of the operating effectiveness of the defined controls; periodic, broadly scoped, and independent testing with expert red-teamers who are industry-renowned and have been recognized in competitive challenges.\\
\textbf{Later (companion document):} Frontier Compliance Framework v2, section 3: Security control monitoring, testing, and assessments: Security controls are regularly tested and independently reviewed to ensure effectiveness. Penetration testing, vulnerability disclosure programs, third party risk assessments, and incident response tabletop exercises aim to help defenses remain robust, and insights from these activities are used to strengthen the company's security posture over time.\\
\textbf{Provider's account:} none found\\
\textbf{Adjudication:} Accepted from first pass (spot-check tier). Re-examination: Changed X to L under rule 11.5: the FCF states that security controls are regularly tested and independently reviewed (all three re-runs).

\paragraph{ANT-2-045 (EC6; X; ANN).} \emph{Anthropic v2.2 \textrightarrow\ v3.0.}\\
\textbf{Earlier:} In the security context, we will delete model weights.\\
\textbf{Later:} NONE\\
\textbf{Provider's account:} [post] Instead, we are choosing to acknowledge these challenges transparently and restructure the RSP before we reach these higher levels. The revised RSP aims to adopt more realistic unilateral commitments that are difficult but still achievable in the current environment, while continuing to comprehensively map the risks we believe the full industry needs to address multilaterally.\\
\textbf{Adjudication:} Weight-deletion commitment removed; post announces removal of hard unilateral commitments as a class, not this one -> ANN-P. Re-examination: Changed ANN-P to ANN, as ANT-2-044.

\paragraph{ANT-2-046 (EC6; W; ANN).} \emph{Anthropic v2.2 \textrightarrow\ v3.0.}\\
\textbf{Earlier:} Monitoring pretraining:We will not train models withcomparable or greater capabilities to the one that requires the ASL-3 Security Standard.13This isachieved by monitoring the capabilities of the model in pretraining and comparing them against the given model. If the pretraining model's capabilities are comparable or greater, we will pause training until we have implemented the ASL-3 Security Standard and established it is sufficient for the model. We will set expectations with internal stakeholders about the potential for such pauses.\\
\textbf{Later:} Anthropic in the lead. We have developed or will imminently develop a highly capable7 model; and we have clear evidence that no other competitor will soon develop such a model. We will require a strong argument that catastrophic risk is contained, along the lines of our recommendations for industry-wide safety (see Section 1). We will delay AI development and deployment as needed to achieve this, until and unless we no longer believe we have a significant lead.\\
\textbf{Provider's account:} [post] Instead, we are choosing to acknowledge these challenges transparently and restructure the RSP before we reach these higher levels. The revised RSP aims to adopt more realistic unilateral commitments that are difficult but still achievable in the current environment, while continuing to comprehensively map the risks we believe the full industry needs to address multilaterally.\\
\textbf{Adjudication:} Agree; removal of the unilateral pause is the announced change. Re-examination: Confirmed.

\paragraph{ANT-2-018 (EC2; W; ANN-P).} \emph{Anthropic v2.2 \textrightarrow\ v3.0.}\\
\textbf{Earlier:} 1. The model is notably more performant on automated tests in risk-relevant domains (defined as 4x or more in Effective Compute3). 2. Six months' worth of finetuning and other capability elicitation methods have accumulated.4This is measured in calendar time, since we do not yet have a metric to estimate the impact of these improvements more precisely.5 In addition, the Responsible Scaling Officer may in their discretion determine that a comprehensive assessment is warranted.\\
\textbf{Later:} NONE\\
\textbf{Provider's account:} none found\\
\textbf{Adjudication:} Accepted from first pass (spot-check tier). Re-examination: Changed X to W and SIL to ANN-P: the Risk Report cadence (every 3 to 6 months) is the counterpart of the comprehensive-assessment trigger (two re-runs); the post states the new cadence but nowhere says the 4x Effective Compute trigger was dropped.

\paragraph{OAI-1-040 (EC4; W; SIL).} \emph{OpenAI beta \textrightarrow\ v2.}\\
\textbf{Earlier:} External access: We will also continue to enable external research and government access for model releases to increase the depth of red-teaming and testing of frontier model capabilities\\
\textbf{Later:} NONE\\
\textbf{Provider's account:} none found\\
\textbf{Adjudication:} Agree; commitment to enable external research and government access removed with no changelog mention. Headline example (EC4). Re-examination: Changed X to W: the discretionary third-party evaluation clause in 5.2 is the weakened counterpart (all three re-runs); no changelog item or post sentence mentions external or government access.

\paragraph{GDM-1-021 (EC2; W; ANN-P).} \emph{Google DeepMind v2.0 \textrightarrow\ v3.0.}\\
\textbf{Earlier:} Machine Learning R\&D uplift level 1: Can or has been used to accelerate AI development, resulting in AI progress substantially accelerating (e.g. 2x) from 2020-2024 rates.\\
\textbf{Later:} ML R\&D acceleration level 1: Has been used to accelerate AI development, resulting in AI progress substantially accelerating from historical rates.\\
\textbf{Provider's account:} none found\\
\textbf{Adjudication:} CCL definition de-specified (quantitative anchor removed; SaferAI Table 10). Post says CCL definitions were "sharpened": class named, direction not this one -> ANN-P.

\paragraph{META-1-032 (EC6; W (MIX); ANN).} \emph{Meta v1.1 \textrightarrow\ v2.}\\
\textbf{Earlier:} Do not release ... Implement mitigations to reduce risk to moderate levels. ... If the results of our evaluations indicate that a frontier AI has a ``high'' risk threshold by providing significant uplift towards realization of a catastrophic outcome we will not release the frontier AI externally.\\
\textbf{Later:} Deploy with mitigations Proceed with deployment of the Frontier AI only if sufficient mitigations are defined, implemented and validated to reduce risk to that of a moderate or lower model. ... In the case where pre-mitigation testing suggests that a model has crossed the high risk threshold, we will not deploy the model externally unless we have strong additional evidence that mitigations are sufficiently robust to adversarial attacks that the fully mitigated model is reduced to moderate or lower levels.\\
\textbf{Provider's account:} [changelog] High threshold measure changed from "Do not release" to "Deploy with mitigations."\\
\textbf{Adjudication:} Agree: consequence for High weakened from do-not-release to deploy-with-mitigations, and the change log states it. Headline example.

\paragraph{MSFT-1-036 (G; W; ANN-P).} \emph{Microsoft v1 \textrightarrow\ feb-2026.}\\
\textbf{Earlier:} Every six months, we will have an explicit discussion on how this framework may need to be improved.\\
\textbf{Later:} At least every twelve months, we will have an explicit discussion on how this framework may need to be improved.\\
\textbf{Provider's account:} [changelog] Adjusting the update cadence to align with regulatory obligations\\
\textbf{Adjudication:} Accepted from first pass (spot-check tier). Re-examination: Changed ANN to ANN-P: the changelog names the update cadence but says only 'adjusting', without direction (all three re-runs).

\paragraph{NAV-1-005 (EC2; W; ANN).} \emph{Naver 2024 \textrightarrow\ v2.0.}\\
\textbf{Earlier:} LLMs should be subject to periodic reviews or assessed whenever major performance improvements are made. ... Our goal is to have AI systems evaluated quarterly to mitigate loss of control risks, but when performance is seen to have increased six times, they will be assessed even before the three-month term is up.\\
\textbf{Later:} NONE\\
\textbf{Provider's account:} [press release] It also replaces a single performance-based criterion with separate criteria for context, use case and impact.\\
\textbf{Adjudication:} Agree; press release states the replacement of the performance-based criterion. Re-examination: Changed X to W: pre-launch impact assessment and post-launch evaluations are the counterpart of the evaluation cadence (two re-runs); the press release states that the performance-based criterion was replaced.

\paragraph{XAI-4-025 (EC2; W; NCL).} \emph{xAI 2025-12-30 \textrightarrow\ 2026-06-30.}\\
\textbf{Earlier:} Thresholds: Our risk acceptance criteria for system deployment is maintaining a dishonesty rate of less than 1 out of 2 on MASK. We plan to add additional thresholds tied to other benchmarks.\\
\textbf{Later:} xAI applies a systemic risk acceptance criteria to each identified risk, incorporating a margin of security, to determine whether each identified systemic risk and the overall systemic risk are acceptable and\\
\textbf{Provider's account:} none found\\
\textbf{Adjudication:} Agree; quantitative MASK criterion becomes qualitative (example F).

\clearpage
\section{Adjudication procedure}
\label{apx:adjudication}
Table~\ref{tab:adjudication} tabulates the coding procedure described in Section~\ref{sec:method}: the first pass, its adjudication, the re-examination of every row that one of the three independent re-runs disputed, and the review of that re-examination against a blind re-coding. The released sheet carries every first-pass code, adjudicated code, re-run code, re-examined code and blind code with its note, and marks for each of the 249 disputed rows whether the blind re-coding agreed, disagreed and was overruled, or disagreed and prevailed. Under the codes that preceded the re-examination the headline figures were a strict rate of 0.67 (257 of 383) and a lenient rate of 0.53, a narrative-against-itemised difference of 0.74 against 0.63 with a permutation $p$ of 0.071, a weakening share of 0.77, and strict silence of 0.75 for weakenings against 0.50 for strengthenings in seven of eight pairs. The re-examination moved 80 rows from silent to partially announced, which lowered the lenient rate and left the strict rate within its interval, and it moved nine rows from partially announced to announced, most of them relocations and consequence removals that the Anthropic 3.0 account describes at class level with their direction. The review then changed 37 rows, namely fifteen announcement codes, nine materiality codes (eight of them rows the re-examination had kept as not material), eight outcome codes and five mixed-direction flags; it moved the lenient rate from 0.32 to 0.30, the strict rate not at all, and the permutation $p$ for the regime difference from 0.23 to 0.13. The regime difference did not survive either step; the direction asymmetry survived both.
\begin{table}[H]\centering\small
\caption{Coding and adjudication counts.}
\label{tab:adjudication}
\begin{tabular}{lr}\toprule Quantity & Value \\ \midrule
Rows in first pass & 710 \\
Rows adjudicated individually & 244 \\
Rows accepted after spot-check & 466 \\
Spot-check sample (accepted rows) & 45 \\
Spot-check changes & 1 \\
First-pass confidence high / medium / low & 271 / 319 / 120 \\
Outcome codes changed by adjudication & 4 \\
Announcement codes changed by adjudication & 56 \\
\quad of which SIL \textrightarrow\ ANN-P & 46 \\
\quad of which ANN \textrightarrow\ ANN-P & 8 \\
\midrule
Rows re-examined after the independent re-runs & 249 \\
\quad materiality codes changed & 19 \\
\quad outcome codes changed & 43 \\
\quad announcement codes changed & 125 \\
\quad\quad SIL \textrightarrow\ ANN-P & 84 \\
\quad\quad ANN-P \textrightarrow\ ANN & 10 \\
\quad\quad SIL \textrightarrow\ ANN & 5 \\
\quad\quad ANN \textrightarrow\ ANN-P & 3 \\
\quad\quad ANN-P \textrightarrow\ SIL & 4 \\
\bottomrule\end{tabular}

\end{table}

\clearpage
\section{Silent revision rate by commitment category}
\label{apx:category}
Table~\ref{tab:cat} supports the category figures cited in Section~\ref{sec:results}. Mat.\ counts all material changes; $n$ counts those on pairs with a revision account. Strict intervals are EC1 (0.30, 0.64), EC2 (0.47, 0.71), EC3 (0.60, 0.82), EC4 (0.36, 0.78), EC5 (0.54, 0.81), EC6 (0.40, 0.65), G (0.66, 0.86), S (0.38, 0.72), M (0.54, 0.83). No test is applied across categories.
\begin{table}[H]\centering\small
\caption{Silent revision rate by commitment category, strict and lenient.}
\label{tab:cat}
\begin{tabular}{llrrrr}
\toprule Code & Category & Mat. & $n$ & $\srr_s$ & $\srr_l$ \\ \midrule
EC1 & Scope of evaluation & 30 & 28 & 0.46 & 0.25 \\
EC2 & Trigger and threshold & 69 & 60 & 0.60 & 0.18 \\
EC3 & Method & 65 & 58 & 0.72 & 0.41 \\
EC4 & Third-party involvement & 20 & 17 & 0.59 & 0.41 \\
EC5 & Disclosure & 47 & 42 & 0.69 & 0.29 \\
EC6 & Consequence of a result & 60 & 57 & 0.53 & 0.28 \\
\midrule
G & Governance & 86 & 67 & 0.78 & 0.36 \\
S & Security & 34 & 29 & 0.55 & 0.24 \\
M & Mitigation & 69 & 34 & 0.71 & 0.29 \\
\bottomrule

\end{tabular}
\end{table}

\clearpage
\section{All labelled version pairs and disclosure regimes}
\label{apx:pairs}
Table~\ref{tab:allpairs} assigns each of the nineteen consecutive labelled pairs in the corpus to a disclosure regime and records whether it was traced, supporting the typology in Section~\ref{sec:corpus}.
\begin{table}[H]\centering\tiny
\caption{Labelled version pairs, regimes and revision-account sources.}
\label{tab:allpairs}
\begin{tabular}{lllllp{4.5cm}}
\toprule Provider & From & To & Regime & Traced & Source of the account \\ \midrule
Anthropic & v1.0 & v2.0 & itemised & yes & In-document 'Changelog' entry 'October 15, 2024 RSP-2024' li \\
Anthropic & v2.0 & v2.1 & itemised & no & RSP page version-history entry 'March 31, 2025' with numbere \\
Anthropic & v2.1 & v2.2 & redline & no & Provider-published redline PDF changelogs/anthropic\_rsp\_v2.2 \\
Anthropic & v2.2 & v3.0 & narrative & yes & Announcement post (anthropic.com/news/responsible-scaling-policy-v3) \\
Anthropic & v3.0 & v3.1 & redline & no & Provider-published redline PDF changelogs/anthropic\_rsp\_v3.1 \\
Anthropic & v3.1 & v3.2 & redline & no & Provider-published redline PDF changelogs/anthropic\_rsp\_v3.2 \\
Anthropic & v3.2 & v3.3 & redline & no & Provider-published redline PDF changelogs/anthropic\_rsp\_v3.3 \\
Anthropic & v3.3 & v3.4 & redline & no & Provider-published redline PDF changelogs/anthropic\_rsp\_v3.4 \\
OpenAI & beta & v2 & itemised & yes & In-document Appendix A 'Change log', twelve numbered items ( \\
Google DeepMind & v1.0 & v2.0 & narrative & no & Announcement post https://deepmind.google/blog/updating-the- \\
Google DeepMind & v2.0 & v3.0 & narrative & yes & Announcement post 'Strengthening our Frontier Safety Framewo \\
Google DeepMind & v3.0 & v3.1 & itemised & yes & In-document section 5.3 'Past Updates and Changes' bullet li \\
xAI & draft-2025-02-10 & draft-2025-02-20 & none & yes & No changelog, version history, post or statement found (chan \\
xAI & draft-2025-02-20 & 2025-08-20 & none & yes & No account found (changelogs/xai\_rmf\_2025-08-20\_2025-08-20\_r \\
xAI & 2025-08-20 & 2025-12-30 & none & yes & No account found (changelogs/xai\_faif\_2025-12-30\_2025-12-30\_ \\
xAI & 2025-12-30 & 2026-06-30 & none & yes & No account found (changelogs/xai\_faif\_2026-06-30\_2026-06-30\_ \\
Meta & v1.1 & v2 & itemised & yes & In-document 'Appendix II - Change log' (v2 PDF p.44) with pe \\
Microsoft & v1 & feb-2026 & itemised & yes & In-document 'Appendix II – Change log' (Feb 2026 PDF p.16).  \\
Naver & 2024 & v2.0 & narrative & yes & ASF 2.0 PDF section '2. The Direction of ASF 2.0' (pp.4-5, p \\
\bottomrule

\end{tabular}
\end{table}

\clearpage
\section{Silent same-label re-uploads}
\label{apx:variants}
Table~\ref{tab:variants} lists the nine files replaced without a new version identifier and the materiality decision for each differing passage, supporting the final paragraph of Section~\ref{sec:results}.
\begin{table}[H]\centering\tiny
\caption{Silent same-label re-uploads and the materiality of each difference.}
\label{tab:variants}
\resizebox{\linewidth}{!}{\begin{tabular}{lp{1.8cm}p{2.4cm}llp{6.6cm}}
\toprule Provider & Base & Variant & Material & Cat. & Change \\ \midrule
Anthropic & v2.0 & v2.0-reupload-20241101 & no & none (changelog text) & Only difference in extracted text: a hyperlink to the v1.0 PDF added to the changelog entry. No commitment text changes. Not material (cross-reference \\
Anthropic & v2.1 & v2.1-reupload-20250402 & no & none (table of contents) & Only difference: the table-of-contents line 'Changelog 17' present in the 1 April file is absent from the 2 April file; the Changelog section itself i \\
OpenAI & v2 & v2-reupload-20250611 & no & none (punctuation) & Typographical changes only (dash glyph, comma after e.g.). Not material. \\
OpenAI & v2 & v2-reupload-20250611 & no & M & The 15 April extraction contains the 'Value alignment' claim twice (an overlaid duplicate text run in the PDF, section C.2 page 19); the 11 June file  \\
Google DeepMind & v2.0 & v2.0-reupload-20250213 & no & none (acknowledgements) & Spelling correction of a contributor's surname. Not material. \\
Google DeepMind & v2.0 & v2.0-reupload-20250328 & no & none (version note) & An 'Updates and changes' page was appended recording a link correction dated 21 March 2025. The corrected hyperlink target is not visible in extracted \\
xAI & 2025-08-20 & 2025-08-20-reupload-20250822 & no & EC3 & The only textual difference: 'AISI' removed from the list of organisations with which the biological-weapons filter topics 'were identified'. The sent \\
Meta & v1.1 & v1.1-reupload-20250328 & yes & EC1 & Scope statement of the whole framework: 'most advanced' and 'match or' deleted, so models that match (rather than exceed) frontier capabilities are no \\
Meta & v1.1 & v1.1-reupload-20250328 & no & EC3 & 'internal deployment' renamed 'closed deployment' in the list of release types the risk assessment considers. Terminology change; the set of release t \\
Meta & v1.1 & v1.1-reupload-20250328 & no & EC1 & Rewording of the same two risk domains; modal force and scope unchanged. Not material. \\
Meta & v1.1 & v1.1-reupload-20250328 & no & EC2 & In the definition of the 'Net new' criterion for a catastrophic outcome, 'i.e.' becomes 'e.g.', turning an exhaustive list of respects (scale, actor,  \\
Meta & v1.1 & v1.1-reupload-20250328 & yes & EC6 & The trigger for the non-release consequence changes its object from 'a catastrophic outcome' to 'a threat scenario'. The document defines threat scena \\
Meta & v1.1 & v1.1-reupload-20250328 & no & none (framing) & Rewording of a framing sentence in section 4.3 Benefits assessment. Not a commitment. Not material. \\
Meta & v1.1 & v1.1-reupload-20250328 & no & none (typography) & Spelling standardisation to US English and font-glyph extraction noise. Not material. \\
Microsoft & v1 & v1-reupload-20250226 & no & none (rendering) & The original rendering had broken text runs ('lor' for 'low or', 'thresh' for 'threshold', a bracketed footnote marker); the re-upload fixes them. Two \\
Magic & v1.0 & v1.0-reupload-20240720 & no & EC2 & The baseline scores that motivate the 50\% LiveCodeBench trigger were replaced with newer models and a different evaluation window; the trigger itself  \\
\bottomrule

\end{tabular}}
\end{table}

\clearpage
\section{Anthropic's Frontier Compliance Framework changelog}
\label{apx:fcf}
The July 2026 document (version 2) carries the changelog in Table~\ref{tab:fcf}, reproduced verbatim from its page 17. Versions 1, 1.1 and 1.2 were not available on the hosting portal at the time of collection. Section~\ref{sec:corpus} treats the withdrawal as a fifth disclosure pattern, and Section~\ref{sec:discussion} reads the document's Section 7.1 as compliance drafting under TFAIA.
\begin{table}[H]\centering\small
\caption{Frontier Compliance Framework changelog, version 2.}
\label{tab:fcf}
\begin{tabular}{llp{9cm}}
\toprule Version & Date & Entry \\ \midrule
v.2 & 24 July 2026 & (i) Revised the Sabotage and Loss of Control Tier 2 (Automated R\&D) threshold in Section 2.4 to align with updates to Anthropic's Responsible Scaling Policy (v3.4) (ii) minor terminology corrections in Section 4. \\
v1.2 & 8 June 2026 & (i) Revised Sabotage and Loss of Control Tier 2 (Automated R\&D) threshold in Section 2.4 to better reflect the underlying threat model and clarify how the threshold is operationalized, (ii) revised our threshold for novel chemical/biological weapons production to better track the threat model of concern; and (iii) made minor terminology changes consistent with updates to Anthropic's Responsible Scaling Policy (v3.1, 3.2). \\
v1.1 & 2 March 2026 & Revised risk tiers in Section 2.4 across all four systemic risk categories to better align with our evolving threat models and capability assessments. Introduced nascent risk tiers for Harmful Manipulation. \\
v.1 & 19 December 2025 & Initial Version \\
\bottomrule
\end{tabular}
\end{table}

\clearpage
\section{Reproducibility statement}
\label{apx:repro}
This appendix states what the release contains, what each reported number depends on, and which parts of the pipeline can be recomputed by a reader. Sections~\ref{sec:corpus} and \ref{sec:method} refer to it.

\paragraph{Release contents.} The repository contains (i) the corpus: every retrieved framework and companion document as published, with a plain-text extraction of each and a SHA-256 hash recomputed from disk; (ii) \texttt{manifest.csv}, one row per document with provider, version label, date, source, retrieval provenance and hash; (iii) the revision accounts, one file per labelled pair, with their source and retrieval date; (iv) the frozen codebook, version 0.2 of 3 September 2026, with its change history; (v) the coding sheet \texttt{tracing\_FINAL\_v3.csv}, 710 rows, carrying the first-pass codes, the adjudicated codes, the codes of the three independent re-runs, the re-examined codes as reviewed, the blind re-coding of the 249 disputed rows with its passages and reasons, every note, the verbatim passages and the changelog pointers; (vi) the three re-run sheets, their run log and the summaries of the stability check and the redlined-pair trace; (vii) the second-coder package, the two returned sheets and the agreement script; and (viii) \texttt{compute\_results.py}, which produces every table, every figure input and every statistic in the paper from (v), with \texttt{make\_figures.py} and \texttt{make\_examples.py}, which draw Figures~\ref{fig:direction} and \ref{fig:srr} and write Appendix~\ref{apx:examples} from the same outputs.

\paragraph{Final codes.} The final code of each variable is the re-examined code where the row was re-examined, with the review's changes written into the same columns and logged in the notes, otherwise the adjudicated code where one was recorded, otherwise the first-pass code. The script applies this precedence to materiality, outcome, announcement status and the flags column alike. A MIX flag set at adjudication and not retained at re-examination is therefore dropped. The \emph{mix} column of Figure~\ref{fig:direction} counts the material changes of each pair whose final flags carry MIX, 54 across the twelve pairs. Running the script with \texttt{--codes adj} stops the precedence at the adjudicated codes and reproduces every number under the codes that preceded the re-examination (Appendix~\ref{apx:adjudication}).

\paragraph{Reproducibility tiers.} Every statistic in Sections~\ref{sec:results} and \ref{sec:discussion} and every appendix table is recomputed from the released coding sheet by the released scripts; a reader who runs them obtains the numbers in the paper exactly. The coding sheet itself is an archived output. A language-model system whose sampling parameters were not fixed assisted with the first pass, and we performed the adjudication and the reviews, so a reader can reproduce the sheet only by re-running the procedure of Section~\ref{sec:method}. A re-run would agree with the released sheet to the degree that the re-runs and the second coding measure, which for announcement status is an $\alpha$ of about 0.85 on all rows and 0.60 to 0.71 on the rows where the three-way choice arises. The corpus and manifest are reproducible in the strongest sense, since every file is hash-pinned and its retrieval source is recorded, and a reader can re-retrieve each provider-hosted document and compare hashes.

\paragraph{Environment.} The scripts require Python 3.10 or later with \texttt{scipy} and \texttt{matplotlib}; no other dependency is used. Figures are drawn by \texttt{make\_figures.py} from the CSV inputs that \texttt{compute\_results.py} writes; the exact permutation test and Wilson intervals are implemented in that script and use no external statistical package beyond \texttt{scipy.stats} for Fisher's exact test and the rank correlations.

\paragraph{Versioning and errata.} The corpus is versioned by release tag. Any correction to a code, a hash or a manifest row is recorded in an errata file in the repository with the date, the affected row and the reason, and the paper's figures are regenerated from the corrected sheet. Provider documents are never altered; a document replaced by its provider is added as a new row and the earlier row is retained.

\clearpage
\section{Corpus manifest}
\label{apx:manifest}
Table~\ref{tab:manifest} lists every row of the released manifest described in Section~\ref{sec:corpus}. C marks companion documents; Acct.\ records whether a provider revision account exists; SHA gives the first eight hexadecimal characters of the SHA-256 hash of the retrieved file.
\begin{table}[H]\centering\tiny
\caption{Corpus manifest.}
\label{tab:manifest}
\resizebox{\linewidth}{!}{\begin{tabular}{lp{3.2cm}llllll}
\toprule Provider & Document & Version & Date & C & Source & Acct. & SHA \\ \midrule
Amazon & Amazon's Frontier Model Safety Fra & 2025-02 & 2025-02-09 &  & provider & na & 0628d781 \\
Anthropic & Anthropic's Responsible Scaling Po & v1.0 & 2023-09-19 &  & provider & na & 14785337 \\
Anthropic & Responsible Scaling Policy & v2.0 & 2024-10-15 &  & provider & yes & cc522e27 \\
Anthropic & Responsible Scaling Policy & v2.0-reupload-20241101 & 2024-10-15 &  & wayback & yes & 22b37ecf \\
Anthropic & Responsible Scaling Policy & v2.1 & 2025-03-31 &  & wayback & yes & c239fc31 \\
Anthropic & Responsible Scaling Policy & v2.1-reupload-20250402 & 2025-03-31 &  & provider & yes & f0ac67ca \\
Anthropic & Responsible Scaling Policy & v2.2 & 2025-05-14 &  & provider & yes & 4807f397 \\
Anthropic & RSP Noncompliance Reporting and An & final-2025-12-04 & 2025-12-04 & C & provider & na & 94f40389 \\
Anthropic & Anthropic Frontier Compliance Fram & v1 & 2025-12-19 & C & not & na & -- \\
Anthropic & Responsible Scaling Policy & v3.0 & 2026-02-24 &  & provider & yes & a71bfa08 \\
Anthropic & Anthropic's Frontier Safety Roadma & feb-2026 & 2026-02-24 & C & wayback & na & bf57607b \\
Anthropic & Anthropic Frontier Compliance Fram & v1.1 & 2026-03-02 & C & not & no & -- \\
Anthropic & RSP Noncompliance Reporting and An & mar-2026 & 2026-03-24 & C & provider & yes & 13eb470a \\
Anthropic & Responsible Scaling Policy & v3.1 & 2026-04-02 &  & provider & yes & 5aa73a3b \\
Anthropic & Responsible Scaling Policy & v3.2 & 2026-04-29 &  & provider & yes & 5410e3d9 \\
Anthropic & Responsible Scaling Policy & v3.3 & 2026-05-26 &  & provider & yes & b7e7cc1e \\
Anthropic & Anthropic Frontier Compliance Fram & v1.2 & 2026-06-08 & C & not & yes & -- \\
Anthropic & Responsible Scaling Policy & v3.4 & 2026-07-08 &  & provider & yes & 6247b9e4 \\
Anthropic & Anthropic Frontier Compliance Fram & v2 & 2026-07-24 & C & provider & yes & 8e4d91e1 \\
Anthropic & Anthropic's Frontier Safety Roadma & jul-2026 & 2026-07-29 & C & provider & yes & 681a531d \\
Cohere & The Cohere Secure AI Frontier Mode & v1.0 & 2025-02-11 &  & provider & na & 9b76fb54 \\
G42 & G42's Frontier AI Safety Framework & 2025-02 & 2025-02-06 &  & wayback & na & 36ddb6b0 \\
Google DeepMind & Frontier Safety Framework & v1.0 & 2024-05-17 &  & provider & na & 3c073cd5 \\
Google DeepMind & Frontier Safety Framework & v2.0 & 2025-02-04 &  & wayback & yes & f82534b5 \\
Google DeepMind & Frontier Safety Framework & v2.0-reupload-20250213 & 2025-02-04 &  & wayback & yes & 5d66eec3 \\
Google DeepMind & Frontier Safety Framework & v2.0-reupload-20250328 & 2025-02-04 &  & provider & yes & 14a48e4e \\
Google DeepMind & Frontier Safety Framework & v3.0 & 2025-09-22 &  & provider & yes & 87ebf40b \\
Google DeepMind & Frontier Safety Framework & v3.1 & 2026-04-17 &  & provider & yes & ad10d2d2 \\
Magic & AGI Readiness Policy & v1.0 & 2024-07-02 &  & wayback & na & 0959ec36 \\
Magic & AGI Readiness Policy & v1.0-reupload-20240720 & 2024-07-02 &  & wayback & no & e7996304 \\
Meta & Frontier AI Framework & v1.1 & 2025-02-03 &  & wayback & na & 9fa301df \\
Meta & Frontier AI Framework & v1.1-reupload-20250328 & 2025-02-03 &  & wayback & no & 8f88ef32 \\
Meta & Advanced AI Scaling Framework & v2 & 2026-04-07 &  & provider & yes & d87aa9bf \\
Microsoft & Frontier Governance Framework & v1 & 2025-02-08 &  & wayback & na & 56f902f1 \\
Microsoft & Frontier Governance Framework & v1-reupload-20250226 & 2025-02-08 &  & provider & no & da0d8b15 \\
Microsoft & Frontier Governance Framework & feb-2026 & 2026-02-01 &  & provider & yes & 3282e4fc \\
NVIDIA & Frontier AI Risk Assessment & 2025-02 & 2025-02-17 &  & provider & na & 8c6aade8 \\
Naver & NAVER's AI Safety Framework (ASF) & 2024 & 2024-06-17 &  & provider & na & 81b60fb7 \\
Naver & NAVER ASF (AI Safety Framework) & 2024-ko-navercorp & 2024-06-17 &  & provider & na & 6aeeefe8 \\
Naver & NAVER AI Safety Framework (ASF) & 2024-ko-clova & 2024-06-17 &  & provider & na & ed5b53e5 \\
Naver & NAVER ASF 2.0 AI Safety Framework  & v2.0 & 2026-07-07 &  & provider & yes & 56e5ee0d \\
Naver & NAVER ASF 2.0 AI Safety Framework  & v2.0-ko & 2026-07-07 &  & provider & yes & 6c86f799 \\
OpenAI & Preparedness Framework (Beta) & beta & 2023-12-18 &  & provider & na & c84e3a59 \\
OpenAI & Preparedness Framework & v2 & 2025-04-15 &  & wayback & yes & 432de80c \\
OpenAI & Preparedness Framework & v2-reupload-20250611 & 2025-04-15 &  & provider & no & fae6e4cc \\
OpenAI & Frontier Governance Framework & 2026-05-28 & 2026-05-28 & C & provider & na & 33e4e118 \\
xAI & xAI Risk Management Framework (Dra & draft-2025-02-10 & 2025-02-10 &  & wayback & na & 5ba8859b \\
xAI & xAI Risk Management Framework (Dra & draft-2025-02-20 & 2025-02-20 &  & provider & no & 89fafebb \\
xAI & xAI Risk Management Framework & 2025-08-20 & 2025-08-20 &  & wayback & no & 31eae94d \\
xAI & xAI Risk Management Framework & 2025-08-20-reupload-20250822 & 2025-08-20 &  & provider & no & 39fab200 \\
xAI & xAI Frontier Artificial Intelligen & 2025-12-30 & 2025-12-30 &  & provider & no & aa01669c \\
xAI & xAI Frontier Artificial Intelligen & 2026-06-30 & 2026-06-30 &  & provider & no & 2c3c6313 \\
\bottomrule

\end{tabular}}
\end{table}

\end{document}